\documentclass[10pt,final,doublecolumn]{IEEEtran}
\IEEEoverridecommandlockouts
\usepackage{amsmath}
\usepackage{amssymb}
\usepackage{latexsym}
\usepackage{graphicx}
\usepackage{bbding}
\usepackage{indentfirst}
\usepackage{cases}
\usepackage{supertabular}
\usepackage{algorithm,algorithmic}
\usepackage{subeqnarray}
\usepackage{color}
\usepackage{bm}
\usepackage{stfloats}
\allowdisplaybreaks[4]

\begin{document}
\title{Cooperative Activity Detection: Sourced and Unsourced Massive Random Access Paradigms}

\author{
Xiaodan Shao, Xiaoming Chen, Derrick Wing Kwan Ng, Caijun Zhong, and Zhaoyang Zhang
\thanks{Part of this paper has been accepted for presentation at the IEEE Global Communications Conference (GLOBECOM), Dec. 2020 \cite{GC}.}
\thanks{Xiaodan Shao ({\tt shaoxiaodan@zju.edu.cn}), Xiaoming Chen ({\tt chen\_xiaoming@zju.edu.cn}), Caijun Zhong ({\tt caijunzhong@zju.edu.cn}), and Zhaoyang Zhang ({\tt ning\_ming@zju.edu.cn}) are with the College of Information Science and Electronic Engineering, Zhejiang University, Hangzhou 310016, China. Derrick Wing Kwan Ng ({\tt w.k.ng@unsw.edu.au}) is with the School of Electrical Engineering and Telecommunications, University of New South Wales, Sydney, NSW 2052, Australia.}}\maketitle

\begin{abstract}
This paper investigates the issue of cooperative activity detection for grant-free random access in the sixth-generation (6G) cell-free wireless networks with sourced and unsourced paradigms. First, we propose a cooperative framework for solving the problem of device activity detection in sourced random access. In particular, multiple access points (APs) cooperatively detect the device activity via exchanging low-dimensional intermediate information with their neighbors. This is enabled by the proposed covariance-based algorithm via exploiting both the sparsity-promoting and similarity-promoting terms of the device state vectors among neighboring APs. A decentralized approximate separating approach is introduced based on the forward-backward splitting strategy for addressing the formulated problem. Then, the proposed activity detection algorithm is adopted as a decoder of cooperative unsourced random access, where the multiple APs cooperatively detect the list of transmitted messages regardless of the identity of the transmitting devices. Finally, we provide sufficient conditions on the step sizes that ensure the convergence of the proposed algorithm in the sense of Bregman divergence. Simulation results show that the proposed algorithm is efficient for addressing both sourced and unsourced massive random access problems, while requires a shorter signature sequence and accommodates a significantly larger number of active devices with a reasonable antenna array size, compared with the state-of-art algorithms.
\end{abstract}


\IEEEpeerreviewmaketitle
\section{Introduction}
Massive machine-type communication (mMTC) or massive access, which is expected to be a typical application scenario for 6G wireless networks, aims to meet the demand of massive connectivity for the Internet-of-Things (IoT) \cite{6g}-\cite{6g2}. Unfortunately, applying conventional grant-based random access schemes for massive access leads to an exceedingly long access latency and a prohibitive signaling overhead. As a remedy, grant-free random access schemes have been proposed and considered as a promising technical candidate for realizing massive access \cite{free}-\cite{kwan5g}, where active devices transmit their data signals without obtaining a grant from their home base stations (BSs) after sending pre-assigned signature sequences.

In general, grant-free random access includes two different paradigms, namely sourced and unsourced random access \cite{amp}-\cite{unsourced}. For sourced random access, the BS is interested in both the messages and the identities of the devices that generated them. In particular, sourced random access is mainly applied to device-oriented applications. For example, in the application of status monitoring, the BS is required to know which sensors send the message and updates their status periodically. Hence, each device is preassigned to a unique signature sequence and the BS identifies the device activity based on the received signals by detecting which sequences are transmitted. Since the active device detection for sourced random access is a typical sparse signal processing problem, numerous compressed sensing (CS)-based approaches have been employed to handle the detection problems. For instance, in \cite{amp}-\cite{amp_gao}, approximate message propagation (AMP) algorithms were designed for joint activity detection and channel estimation (JADCE) in different scenarios by exploiting the statistics of wireless channels. Then, the authors in \cite{oamp} proposed an orthogonal AMP-multiple measurement vectors (OAMP-MMV) algorithm that applies Hadamard pilot matrice and reduces the computational complexity by using fast Fourier transform. In addition, the authors in \cite{shaodim} proposed a low-complexity dimension reduction-based JADCE algorithm, which projects the original device state matrix to a low-dimensional space by exploiting its sparse and low-rank structure. Note that the aforementioned approaches in \cite{amp}-\cite{shaodim} performed activity detection based on the instantaneous received signals, which require exceedingly long signature sequences in the scenario of massive access. To tackle this problem, the covariance-based algorithms were proposed to improve the performance of device activity detection in \cite{covar} and \cite{covar1}, where the detection problem was handled by a coordinate descent algorithm. After that, a low computational complexity covariance-based algorithm with constant-modulus pilots was proposed for JADCE, which shows low computational complexity and appealing activity detection performance \cite{cheng}. In fact, the covariance-based algorithm can outperform the AMP algorithm with the same length of signature. However, the superior system performance comes from the expense of the use of a relatively large number of receive antennas compared with what traditional CS-based algorithms needed. As a result, active device detection in massive access has emerged as a challenging problem due to a large number of devices and the limited radio resources in $6$G wireless networks.

As for unsourced random access, the AP is interested in the transmitted messages only and not the identity of the devices. In fact, unsourced random access is motivated by practical IoT scenarios, where millions of low-cost devices have their codebook hardwired at the moment of production \cite{concatenated}. Such that all the devices share a common codebook in unsourced random access. Since the same codebook is exploited, the AP can only decode the list of transmitted messages irrespectively of the identity of the active devices. In this context, the computational complexity can be significantly reduced compared with the case that assigns a unique individual codebook to each device \cite{free}. Therefore, unsourced random access is mainly applied to content-oriented applications. For example, in quality inspections process of smart factories, a fraction of devices send their current reliability information to the AP, while the AP detects all received messages, computes a weighted average of current informations, then generates a performance index. Recently, a low-complexity unsourced random access algorithm based on the coupled CS framework was proposed in \cite{ura-single-ante}. Specifically, the transmission slot is partitioned into subslots and each active device sends a codeword from a common codebook across different subslots. To this connection, an inner encoder is needed which maps each submessage into one column of a given coding matrix. Then, at the BS, the inner decoder must identify which columns of the matrix have been transmitted. On the other hand, an outer tree-based decoder is applied to stitch the decoded sequences. However, this work assumed only a single-antenna receiver equipped at the BS and the results are not applicable to the case of multiple antennas. Afterward, the authors in \cite{unsourced} extended the model in \cite{ura-single-ante} to a case of large-scale receive antenna arrays with Rayleigh fading, where a maximum likelihood (ML)-based activity detection scheme was adopted as the inner decoder of a concatenated coding scheme in \cite{concatenated}. Specifically, the algorithm in \cite{unsourced} avoided the use of a signature sequence longer than the number of active devices. However, it usually requires a large number of antennas at the BS in order to accumulate more active devices \cite{kwanantenna}. Moreover, for a short subslot length, the number of active devices that can be accommodated is limited by the detection capabilities of the inner decoder \cite{unsourced}.

To overcome these challenges in sourced and unsourced random access, cooperative activity detection with multiple APs can be applied to detect the active device and the active columns of the coding matrix. Inspired by this, the authors in \cite{multicell} considered the active device detection in multi-cell massive multiple-input-multiple-output (MIMO) systems and cooperative MIMO systems using the AMP algorithm. For multi-cell massive MIMO, the authors assumed that each AP operates independently to detect the active devices of its own cell. In this case, inter-cell interference is treated as noise, which is a severe limiting factor for achieving reliable activity detection. As for cooperative MIMO, each AP performs active device detection locally and detects the devices from neighboring cells as well. Then, the detection results in the form of log-likelihood ratio are forwarded to a central unit where final decisions on device activity are carried out. Then, the authors in \cite{multicell_rm} proposed an active device identification method in a multi-cell network adopting second-order Reed-Muller sequences, where each AP identifies all in-cell devices and the transmissions from out-of-the-cell devices are regarded as interference. Compared with \cite{multicell} and \cite{multicell_rm}, the proposed cell-free framework in this paper can avoid the inter-cell interference due to the fact that the concepts of cell and cell boundary do not exist in our system.

Motivated by these works, this paper designs a unified framework for both sourced and unsourced activity detection in 6G cell-free wireless networks \cite{cellfree}-\cite{cellfree2}, where multiple APs are deployed in a vast area to serve all devices located in this area via a fronthaul network \cite{fronthaul}. For instance, the authors in \cite{cellfree2} studied grant-free massive access in cell-free massive MIMO based IoT, where multiple APs cooperate in the network to serve massive devices. Herein, the connections between APs can be set differently according to the communication radius. Unlike the conventional cooperative MIMO where the final decisions on device activity are only carried out in a central unit, the proposed framework does not need a centralized fusion center and the cooperative activity detection among the APs only needs to exchange low-dimensional intermediate information, i.e., device state vectors, thereby all the APs in the system can obtain the detection results. Therefore, it is reliable and robust to AP and/or fronthaul link failure. More importantly, to reduce the amount of associated signaling overhead under this framework, this paper designs a scalable computationally efficient algorithm to detect the activity. In summary, our main contributions are listed as follows:

\begin{enumerate}
\item We propose a unified cooperative activity detection framework for sourced and unsourced random access based on the covariance of the received signals in 6G cell-free wireless networks to support massive IoT with heterogeneous application requirements.

\item We develop a novel low-complexity cooperative activity detection (CAD) algorithm that exploits the special characteristic of the device state vectors of interest among neighboring APs, namely joint similarity and sparsity.

\item We provide a theoretical analysis of the convergence property of the proposed CAD algorithm and the result shows that it enjoys a convergence rate of $\mathcal{O}(1/t)$ under certain general conditions with an iteration index $t$. Moreover, extensive simulation results confirm the effectiveness of the proposed CAD algorithm in both sourced and unsourced random access paradigms.
\end{enumerate}

The rest of this paper is organized as follows. Section II gives a brief introduction of 6G cell-free wireless networks in a sporadic device activity pattern scenario. Then, Section III proposes the cooperative sourced random access scheme. Section IV designs the cooperative unsourced random access scheme. Next, Section V analyzes the performance of the proposed CAD algorithm. Afterward, Section VI provides extensive simulation results to illustrate the performance of the proposed algorithm. Finally, Section VII concludes the paper.

\emph{Notations}: We use bold letters to denote matrices or vectors, non-bold letters to denote scalars, $\mathbf{I}$ to denote the identity matrix, $\mathbb{C}^{A\times B}$ to denote the space of complex matrices of size $A\times B$, $|\cdot|$ to denote the absolute value of a complex number, $(\cdot)^H$ and $(\cdot)^T$ to denote conjugate transpose and transpose respectively, $\left\|\cdot\right\|_F$ to denote Frobenius norm of a matrix, $\left \| \cdot \right \|_{2}$ to denote the $l_2$-norm of an input vector. $\|\cdot\|_{0}$ denotes the $l_{0}$-norm defined as the number of nonzero elements of an input vector. $\mathbf{A}(n,:)$ denotes
the $n$th row of matrix $\mathbf{A}$. $\|\mathbf{A}\|_{2,0}$ denotes the $l_{20}$-norm defined as the number of nonzero elements of vector $\left[\|\mathbf{A}(1,:)\|_2, \cdots,\|\mathbf{A}(n,:)\|_2\right]$. $\det(\cdot)$ and $\text{tr}(\cdot)$ are operators that return the determinant and the trace of an input matrix, respectively. $\left \langle \mathbf{a},\mathbf{b} \right \rangle$ denotes the inner product of the vectors $\mathbf{a}$ and $\mathbf{b}$. $\text{col}(\cdot)$ denotes a column vector. $|\cdot|_c$ denotes the cardinality of a set. $\odot$ denotes element wise multiplication. $\bigtriangledown f(\cdot)$ denotes the gradient of a function $f(\cdot)$. $\partial f(\cdot)$ denotes the subgradient of a function $f(\cdot)$. $\mathcal{O}(\cdot)$ stands for the big-O notation.

\section{System Model}
\begin{figure}
  \centering
\includegraphics [width=0.46\textwidth] {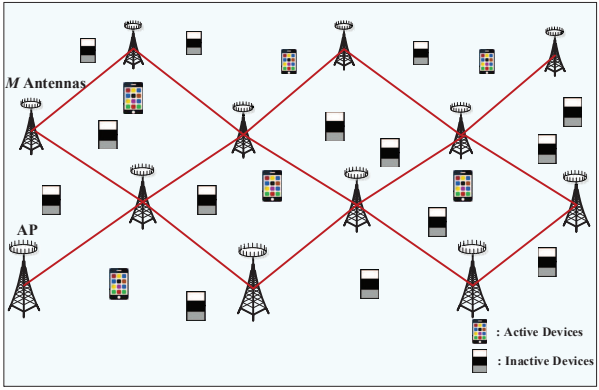}
\caption{Illustration of a 6G cell-free wireless network with multiple APs.}
\label{cooperative}
\end{figure}

Consider a 6G cell-free wireless network comprising $B$ APs. The APs are equipped with $M$ antennas each, serving $N$ uniformly distributed single-antenna IoT devices in a vast area. Each AP is connected to several adjacent APs via fronthaul links and can only communicate with its one-hop neighbors for reducing the communication load, as shown in Fig. \ref{cooperative}. In 6G wireless networks, the density of IoT devices is usually huge, e.g., $10$ devices per $m^2$. However, due to the bursty communication characteristic of IoT applications, only a fraction of IoT devices are active at any given time slot. We adopt $\mathcal{K}$ to denote the set of active devices with cardinality $K=\left|\mathcal{K}\right|_c\ll N$ being the number of the active devices, which is a random variable. For convenience, we define $\chi_n$ as a binary activity indicator with ${\chi_n} = 1$ if the $n$th device is active, and ${\chi_n} = 0$ otherwise. Moreover, we represent the $M$-dimensional channel vector from the $n$th device to the $b$th AP as $\sqrt{g_{b,n}}\mathbf{h}_{b,n}$, where $g_{b,n}$ is the large-scale fading depending on the location of the $n$th device and $\mathbf{h}_{b,n} \in \mathbb{C}^{M}$ is the corresponding small-scale fading following independent and identically distributed (i.i.d.) complex Gaussian distribution with zero mean and unit variance.

In order to reduce the access latency and the system signaling overhead, a grant-free random access protocol is advocated for 6G cell-free wireless networks \cite{free}. Specifically, at the beginning of each time slot, the active devices transmit the corresponding signature sequences over uplink channels simultaneously and then the APs perform the activity detection based on the received signal cooperatively. Herein, it is assumed that all signature sequences $\mathbf{s}_n \in \mathbb{C}^{L}$ are generated following i.i.d. complex Gaussian distribution with zero mean and unit variance which are known at the APs in advance. Thus, the received signal $\mathbf{Y}_b\in \mathbb{C}^{L\times M}$ at the $b$th AP can be expressed as
\begin{eqnarray}
\label{aps}
\mathbf{Y}_b=\sum_{n=1}^N\chi_n\mathbf{s}_n \sqrt{g_{b,n}}\mathbf{h}_{b,n}^T+\mathbf{W}_b=\mathbf{S}\boldsymbol{\Gamma}_b^{\frac{1}{2}}\mathbf{H}_b+\mathbf{W}_b,
\end{eqnarray}
where $\mathbf{H}_b=[\mathbf{h}_{b,1},\cdots,\mathbf{h}_{b,N}]^T\in \mathbb{C}^{N \times M}$ denotes the small-scale fading channel matrix, $\mathbf{S}=[\mathbf{s}_1,\cdots,\mathbf{s}_N]\in \mathbb{C}^{L \times N}$ denotes the horizontal stack of all signature sequences, and $\mathbf{W}_b \in \mathbb{C}^{L \times M}$ is the additive white Gaussian noise (AWGN) marix with i.i.d. entries $\thicksim\mathcal{CN}(0,\sigma^2)$, where $\sigma^2$ denotes the noise power at each antenna. Define $\boldsymbol{\gamma}_b =[\gamma_{b,1},\cdots,\gamma_{b,N}]^T\in \mathbb{R}^{N}$ as the diagonal entries of diagonal matrix $\boldsymbol{\Gamma}_b$, representing the device state vector of the $b$th AP with $\gamma_{b,n}=\chi_n g_{b,n}$.

In this paper, we aim to provide two grant-free random access schemes according to the characteristics and requirements of IoT applications. The first kind of IoT applications have to know which device sends the messages. In this case, we should detect the identity of active device, namely sourced random access. The second kind of IoT applications only need to know what messages are sent. In this case, we directly detect the messages, namely unsourced random access. In the following, we will design sourced and unsourced random access schemes via exchanging some intermediate variables based on a limited cooperation among multiple APs.

\section{Cooperative Sourced Random Access}
For sourced random access, each device is assigned to an unique sequence. Based on the received sequences, the APs perform the cooperative device activity detection. In this section, we first propose a cooperative detection framework for 6G cell-free wireless networks with a massive number of IoT devices. Then, we design a corresponding cooperative detection algorithm for such a sourced random access scenario.

\subsection{Cooperative Massive Detection Framework}
For the activity detection problem based on model in \eqref{aps}, the unknown device state vectors for different APs are generally different. Moreover, there are some common characteristics among the neighboring APs. To enhance the detection
performance, we first associate a local estimator for each AP due to the fact that different AP estimates a different device state vector. Then, to incorporate the estimates of neighboring APs, i.e., sparsity-promoting and the similarity-promoting terms \cite{shaodim, tls}, we can modify the local estimator to associate a regularized local cost function with each AP.

\subsubsection{Covariance Based Local Estimator}
Firstly, we design the local estimator. It is well known that the covariance-based massive activity detection is equivalent to recovering the device state vector $\boldsymbol{\gamma}_b$ from the noisy measures $\mathbf{Y}_b$ with the knowledge of the pre-defined sequence matrix $\mathbf{S}$.
In general, the estimation of the device state vector $\boldsymbol{\gamma}_b$ can be formulated as a ML estimation problem \cite{covar}. In particular, for a given $\mathbf{Y}_b$, each column of $\mathbf{Y}_b$, denoted as $\mathbf{y}_{bm}$, $1 \leq m \leq M$, can be termed as an independent sample having the following multivariate complex Gaussian distribution:
\begin{equation}\label{ml}
\mathbf{y}_{bm}\sim \mathcal{CN}(\mathbf{0},\mathbf{S}\boldsymbol{\Gamma}_b\mathbf{S}^H+\sigma^2\mathbf{I}),
\end{equation}
where covariance matrix is calculated by $\mathbb{E}\left \{ \mathbf{y}_{bm}\mathbf{y}_{bm}^H\right\}$. For convenience, we define
$\boldsymbol{\Sigma}_b=\mathbf{S}\boldsymbol{\Gamma}_b\mathbf{S}^H+\sigma^2\mathbf{I}$.
Then, the likelihood of $\mathbf{Y}_b$ given $\boldsymbol{\gamma}_b$ can be represented as
\begin{eqnarray}\label{liki}
P(\mathbf{Y}_b|\boldsymbol{\gamma}_b)
=\frac{1}{\det(\pi\boldsymbol{\Sigma}_b)^M}\exp(-\text{tr}(\boldsymbol{\Sigma}_b^{-1}\mathbf{Y}_b\mathbf{Y}_b^H)).
\end{eqnarray}
By exploiting the Gaussianity, we can obtain the ML estimator of $\boldsymbol{\gamma}_b$ at the $b$th AP as follows:
\begin{eqnarray}\label{eqr}
  f(\boldsymbol{\gamma}_b)=-P(\mathbf{Y}_b|\boldsymbol{\gamma}_b)=\ln\text{det}(\boldsymbol{\Sigma}_b)+\text{tr}(\boldsymbol{\Sigma}_b^{-1}\hat{\boldsymbol{\Sigma}}_{b\mathbf{y}}),
\end{eqnarray}
where $\hat{\boldsymbol{\Sigma}}_{b\mathbf{y}}=\frac{1}{M}\mathbf{Y}_b\mathbf{Y}_b^H$ denotes the sample covariance matrix of the received signal of the $b$th AP averaged over different antennas. Based on \eqref{eqr}, the ML estimation problem can be formulated as $\arg \min_{\boldsymbol{\gamma}_b\in\mathbb{R}_+}f(\boldsymbol{\gamma}_b)$.

\subsubsection{Sparsity-Promoting Term Design}
Secondly, since the focused device state vectors have inherently structured sparsity, we propose a sparsity-promoting term to facilitate cooperative detection among multiple APs. In \cite{shaodim}, the rows of the interested matrix share the common support, thus appropriate co-regularizer was added to the mean-square error criterion to promote row sparsity. In this paper, the specific sparsity pattern can be simultaneously observed at different APs, namely the indices of nonzero entries of $\boldsymbol{\gamma}_b$ are consistent for $b=1,2,\cdots,B$. Because each AP only communicates with its neighboring APs, it cannot obtain the global information of the system-level sparsity pattern. Moreover, it is quite challenging to split this global quantity into several local quantities consisting of components only from the neighboring nodes. In this case, for the $b$th AP, we define a local parameter matrix consisting of the parameter vectors of all its one-hop neighbors, which can be directly obtained as follows:
\begin{align}\label{spas1}
  \mathbf{R}_b\!=\!\left [ \boldsymbol{\gamma}_{l_1},\boldsymbol{\gamma}_{l_2},\boldsymbol{\gamma}_{l_i},\cdots,\boldsymbol{\gamma}_{l_{|\mathcal{N}_b^{-}|_c}},\boldsymbol{\gamma}_b \right ]&  \in \mathbb{C}^{N \times (|\mathcal{N}_b|_c)},\nonumber\\
  & \forall b\in \{1,2,\cdots,B\},
\end{align}
where $l_i\in \mathcal{N}_b^{-}$ is the index set of neighbors of the $b$th AP except itself, $\mathcal{N}_b$ denotes the index set of the neighbors of the $b$th AP including itself. Consequently, we aim to promote row sparsity of matrix $\mathbf{R}_b$ to exploit the joint sparsity. In general, the $l_{20}$-norm which is a combination of $l_0$-norm and $l_2$-norm can promote the row sparsity of an input matrix \cite{boy}. However, the $l_{20}$-norm
is non-convex and non-differentiable, and its minimization is non-deterministic polynomial-time hardness. To this end, this paper adopts the following sparsity-promoting term proposed in our work \cite{shaodim}
\begin{equation}\label{spas}
   g(\boldsymbol{\gamma}_b)= \sum_{n=1}^{N}\underbrace{\left(\left \| \mathbf{R}_b(n,:) \right \|_2-\frac{1}{\theta}\ln(1+\theta\left \| \mathbf{R}_b(n,:) \right \|_2)\right)}_{\Phi(\mathbf{R}_b(n,:))},
\end{equation}
Calculating the Maclaurin series of $\ln(1+\theta\left \| \mathbf{R}_b(n,:)\right \|_2)$, we note that when $\|\mathbf{R}_b(n,:)\|_2 \rightarrow0$, $\Phi(\mathbf{R}_b(n,:))\rightarrow \frac{\theta}{2}\|\mathbf{R}_b(n,:)\|_2^2$, however, when $\|\mathbf{R}_b(n,:)\|_2 \rightarrow\infty$, $\Phi(\mathbf{R}_b(n,:))\rightarrow \|\mathbf{R}_b(n,:)\|_2$. Herein, $\theta > 0$ is the penalty parameter used to guarantee that the curve of $\Phi(\mathbf{R}_b(n,:))$ is as close as possible to the curve of $\|\mathbf{R}_b(n,:)\|_2$. In other words, the function $\Phi(\mathbf{R}_b(n,:))$ is used to approximates the $l_2$-norm of $\mathbf{R}_b(n,:)$, since $\Phi(\mathbf{R}_b(n,:))$ is differentiable with respect to the row vector $\mathbf{R}_b(n,:)$ at $\mathbf{0}$.
Moreover, the $g(\boldsymbol{\gamma}_b)$ is convex since it relaxes the non-convex $l_0$-norm by the summation operator. The nonzero rows are penalized by minimizing $g(\boldsymbol{\gamma}_b)$. In this way, a common sparsity profile across the columns of the local parameter matrix $\mathbf{R}_b$ is promoted. Although the sparsity-promoting term is imposed on the local parameter matrix $\mathbf{R}_b$, the cooperative nature promotes a common sparsity profile across all columns of the global device state vectors $\{\boldsymbol{\gamma}_b\}_{b=1}^B$. Note that depends on the assumptions imposed on the problem, one may design other appropriate penalties for sparsity-promoting term containing smooth or non-smooth functions.

\subsubsection{Similarity-Promoting Term Design}
Thirdly, we design a similarity-promoting term to improve the detection performance. In \cite{spars} and \cite{tls}, to promote similarity among optimum parameter vectors, a $l_2$-norm based co-regularizer has been introduced to the diffusion least-mean-square algorithm. In the considered problem, the device state vectors of interest among neighboring APs also exhibit similarity. Specifically, the indices of nonzero entries of the global device state vector $\{\boldsymbol{\gamma}_b\}_{b=1}^B$ for all APs should be the same, but the amplitudes of the nonzero entries at the APs are different from each other due to the effects of different path losses. In particular, since the neighboring APs have short distances, the corresponding device state vectors consist of some similar nonzero entries.
Moreover, the ML estimator $f(\boldsymbol{\gamma}_b)$ depends on the empirical covariance $\hat{\boldsymbol{\Sigma}}_{b\mathbf{y}}$. In high-dimensional settings, where
the length of sequences $L$ is larger than the number of AP antennas $M$, $\hat{\boldsymbol{\Sigma}}_{b\mathbf{y}}$ will be relatively different from the covariance matrix $\boldsymbol{\Sigma}_b$. By enforcing structural similarity, the update of each $\boldsymbol{\Sigma}_b$ can exploit from the fact that the estimates of neighboring APs should be similar to each other.
Motivated by these observations, we design a similarity-promoting function as follows
\begin{equation}\label{sim}
\Psi(\boldsymbol{\gamma}_b)=\sum_{l\in \mathcal{N}_b}c_{lb}\Psi_l(\boldsymbol{\gamma}_b-\boldsymbol{\gamma}_l),~\forall b\in \{1,2,\cdots,B\},
\end{equation}
where $\Psi_l(\boldsymbol{\gamma}_b-\boldsymbol{\gamma}_l)$ is a convex penalty function, minimized at $\Psi_l(\mathbf{0})$, which encourages similarity between $\boldsymbol{\gamma}_b$ and $\boldsymbol{\gamma}_l$. In this paper, the specific expression in the penalty function $\Psi_l(\boldsymbol{\gamma}_b-\boldsymbol{\gamma}_l)$ form is set to $l_1$-norm penalty $\Psi_l(\mathbf{x})=\sum_{n=1}^N|x_n|$, where $x_n$ denotes the $n$th element of the vector $\mathbf{x}$. This penalty function can encourage a large number of elements of device state vector to be identical across APs. In other words, it borrows information aggressively across neighbors, encouraging not only similar structure but also similar values. The adopted penalty is suitable for massive access where only a small fraction of potential devices are active at a time slot. Herein, $c_{lb}$ are linear weights satisfying the conditions:
\begin{equation}\label{clbo}
  \sum\limits_{l \in {\mathcal{N}_b}} {c_{lb}} = 1, ~\textrm{and}~~{c_{lb}} = 0~~ \forall l \notin {\mathcal{N}_b}.
\end{equation}

After defining the similarity-promoting term and sparsity-promoting term, the amalgamation of \eqref{eqr}, \eqref{spas}, and \eqref{sim} leads to the following novel regularized local cost function to be adopted at the $b$th AP:
\begin{eqnarray}\label{ob1}
F(\boldsymbol{\gamma}_b)\!=\!f(\boldsymbol{\gamma}_b)
+ \beta g(\boldsymbol{\gamma}_b)+\tau\Psi(\boldsymbol{\gamma}_b), ~\forall b\!\in\! \{1,2,\cdots,B\},
\end{eqnarray}
where $\beta>0$ and $\tau>0$ are the penalty parameters used to enforce sparsity and similarity, respectively \cite{Jalali}. By varying the values of these penalty parameters, one can strike a balance between the ML estimator and promoting terms. In \eqref{ob1}, in addition to exploiting the local received signal through certain covariance matrix, the structured sparsity and similarity pattern observed in neighboring APs are leveraged to improve activity detection performance. To ensure that each AP can solve the problem in an autonomous and adaptive manner using only local interactions, in Section III. B and Section III. C, we will design a cooperative activity detection (CAD) algorithm to jointly minimize the costs involved \eqref{ob1} at each AP.

\subsection{A Decentralized Approximate Separating Strategy}
Note that the first term of \eqref{ob1}, i.e., $f(\boldsymbol{\gamma}_b)$ is differentiable and geodesically convex, which is a generalized form of classical convexity and can guarantee that all local minima of geodesic convexity functions are globally minimum \cite{geo}. However, as stated in the above subsection, $\Psi_l(\boldsymbol{\gamma}_b-\boldsymbol{\gamma}_l)$ is discontinuous, i.e., the third term of the local cost function could be a sum of non-smooth functions, and the second term is also potentially non-differentiable. In addition, the unknown variables $\boldsymbol{\gamma}_b$ for neighboring APs are coupled with each other. These obstacles make the problem intractable to solve and existing algorithms, e.g., \cite{notsolve} and \cite{tls}, are not applicable to such a problem.
In this paper, we aim to design a scheme which computes device state vectors of all APs independently to facilitate decentralized implementation. In the following, we design a decentralized approximate separating strategy for minimizing the cost function in \eqref{ob1} based on the forward-backward splitting strategy \cite{forward}, which can handle the non-smooth problem and is especially amenable to solve the high-dimensional activity detection problem due to its fast convergence rate and its conceptual and mathematical simplicity.

Before proceeding, we recall the forward-backward splitting approach for minimizing \eqref{ob1}, which is given by the iteration
\begin{eqnarray}\label{gbp}
  \boldsymbol{\gamma}_b^{t+1}&=&\boldsymbol{\gamma}_b^{t}-\eta_b^t\bigtriangledown f(\boldsymbol{\gamma}_b^t)-\tau\eta_b^t\partial\Psi(\boldsymbol{\gamma}_b^{t+1})-\beta\eta_b^t\partial g(\boldsymbol{\gamma}_b^{t+1}) \nonumber \\
&=&\overbrace{\text{prox}_{\eta_b^t(\tau\Psi+\beta g)}(\underbrace{\boldsymbol{\gamma}_b^{t}-\eta_b^t\bigtriangledown f(\boldsymbol{\gamma}_b^t)}_{\textrm{Forward step}})}^{\textrm{Backward step}},
\end{eqnarray}
where $\eta_b^t$ is the step size for the $b$th AP at iteration $t$, and $\boldsymbol{\gamma}_b^t$ denotes the value of $\boldsymbol{\gamma}_b$ in the $t$th iteration.
Note that $\Psi(\boldsymbol{\gamma}_b^{t+1})$ and $\partial g(\boldsymbol{\gamma}_b^{t+1})$ have the effect of evaluating the gradient at $\boldsymbol{\gamma}^{t+1}$. This form is known to have better approximation properties than conventional gradient descent algorithm which evaluates the gradient at $\boldsymbol{\gamma}^{t}$ \cite{forward}.
However, it cannot be rewritten as an iteration that gives $\boldsymbol{\gamma}_b^{t+1}$ explicitly in terms of $\boldsymbol{\gamma}_b^{t}$. For this reason, the proximal operator of a function $h$ is introduced, which is is a mapping function given by:
\begin{equation}\label{pr}
  \text{prox}_{\eta h}(\mathbf{y})=\arg\min_{\mathbf{u}} h(\mathbf{u})+\frac{1}{2\eta}\left \| \mathbf{u}-\mathbf{y} \right \|_2^2,
\end{equation}
with variables $\mathbf{y}$ and $\mathbf{u}$, and a step-size $\eta > 0$ \cite{acce}. As shown in the last equation of \eqref{gbp}, the gradient descent step is the forward step and the proximal step is the backward step. Herein, the proximal step can be viewed as an implicit discretization of the non-differentiable function, which can be applied to overcome the non-smoothness of $\Psi(\boldsymbol{\gamma}_b)$ and the proximal algorithm has a faster convergence rate than the approach based on the subgradient strategy \cite{forward}.

Unfortunately, it is prohibitively challenging to directly evaluate the proximal operators with respect to similarity-promoting function $\Psi(\boldsymbol{\gamma}_b)$ and the sum of $\beta g(\boldsymbol{\gamma}_b) + \tau\Psi(\boldsymbol{\gamma}_b)$. Moreover, the calculation of $\Psi(\boldsymbol{\gamma}_b)$ over all the number of neighborhood, $|\mathcal{N}_b|_c$, in each iteration is expensive. Motivated by Douglas Rachford splitting in \cite{forward}, where two of the proximal operators can be updated alternately, this paper aims to handle the proximal operator of function $\Psi(\boldsymbol{\gamma}_b)$ and $g(\boldsymbol{\gamma}_b)$ separately. Specifically, we first calculate an estimator $\mathbf{x}_b^t$ of the subgradient $\partial\Psi(\boldsymbol{\gamma}_b^t)$ and then incorporate the gradient descent step into the proximal step with respect to sparsity-promoting term $g(\cdot)$ for the $b$th AP, which is given by
\begin{eqnarray}\label{zt}
\mathbf{z}_b^{t}
  &=&\boldsymbol{\gamma}_b^t-\eta_b^t\bigtriangledown f(\boldsymbol{\gamma}_b^t)-\tau\eta_b^t\mathbf{x}_b^t-\beta\eta_b^t\partial g(\mathbf{z}_b^{t}) \nonumber\\
  &=&\text{prox}_{\beta\eta_b^t g}(\boldsymbol{\gamma}_b^t-\eta_b^t\bigtriangledown f(\boldsymbol{\gamma}_b^t)-\tau\eta_b^t\mathbf{x}_b^t),
\end{eqnarray}
where $\mathbf{z}_b^{t}$ is an intermediate variable. Then, according to the update rule of Douglas Rachford splitting, we incorporate $\mathbf{z}_b^{t}$ into proximal operator with respect to similarity-promoting function $\Psi(\boldsymbol{\gamma}_b)$:
\begin{eqnarray}\label{DR}
  \boldsymbol{\gamma}_b^{t+1}=\text{prox}_{\tau\eta_b^{l,t} \Psi}(\mathbf{z}_b^t+\tau\eta_b^{t} \mathbf{x}_b^{t}).
\end{eqnarray}

The intermediate variable $\mathbf{z}_b^{t}$ and device state vector $\boldsymbol{\gamma}_b^{t}$ iterate alternately and their values approach to each other. When converging to optimality, their values are identical. In order to overcome the difficulty in processing the non-smooth finite sum term and to reduce the computational overhead of \eqref{DR}, this paper proposes a decentralized separating strategy that approximately separates the proximal operator of the similarity-term of the $b$th AP, $\Psi(\boldsymbol{\gamma}_b)$, into the proximal operator of one of the neighbors of AP $b$, $\Psi_l(\boldsymbol{\gamma}_b-\boldsymbol{\gamma}_l)$, in each iteration. In mathematical terms, we first choose $l$ randomly from the set $\mathcal{N}_b$ with probabilities $\{p_1,p_2,\cdots,p_{\left | \mathcal{N}_b \right |_c}\}$. Then, we incorporate $\mathbf{z}_b^{t}$ into proximal operator with respect to the sub-function of  similarity-promoting $\Psi_l(\boldsymbol{\gamma}_b-\boldsymbol{\gamma}_l)$:
\begin{eqnarray}\label{mainstep}
  \boldsymbol{\gamma}_b^{t+1}&=&\mathbf{z}_b^{t}-\tau\eta_b^{l,t}\left[\partial\Psi_l(\boldsymbol{\gamma}_b^{t+1})-\mathbf{x}_b^{l,t}\right]\nonumber\\
  &=&\text{prox}_{\tau\eta_b^{l,t} \Psi_l}(\mathbf{z}_b^t+\tau\eta_b^{l,t} \mathbf{x}_b^{l,t}),
\end{eqnarray}
where $\mathbf{x}_b^{l,t}$ is the estimator of subgradient $\partial\Psi_l(\boldsymbol{\gamma}_b^{t+1})$ for the randomly selected $l$th neighbor of the $b$th AP in the $t$th iteration. Let
$c_{lb}^t$ denote the combiner at $t$th iteration and set $\eta_b^{l,t}=\frac{c_{lb}^t\eta_b^t}{ p_l}$, which is a stochastic approximation of $\eta_b^t$ controlled by the combiner and the probability of being selected. In this manner, we are able to treat the difficult terms in \eqref{ob1} with non-smooth finite sum term for any size of cardinality $|\mathcal{N}_b|_c$.

Since $\mathbf{z}_b^{t}$ and $\boldsymbol{\gamma}_b^{t}$ converge to the same value,
\eqref{mainstep} is an accurate approximation of \eqref{gbp} if $\mathbf{x}_b^{l,t}=\partial\Psi_l(\boldsymbol{\gamma}_b^{t+1})$ and $\mathbf{x}_b^{t}=\partial\Psi(\boldsymbol{\gamma}_b^{t+1})$ hold. Thus, we must ensure that $\partial\Psi_l(\boldsymbol{\gamma}_b^{t+1})$ is close to $\mathbf{x}_b^{l,t}$ to obtain an accurate estimator. According to the definition of proximal operator in \eqref{pr}, equation \eqref{mainstep} satisfies
\begin{align}\label{xwl}
\frac{(\mathbf{z}_b^t\!+\!\tau\eta_b^{l,t} \mathbf{x}_b^{l,t}\!-\!\text{prox}_{\tau\eta_b^{l,t} \Psi_l}(\mathbf{z}_b^t\!+\!\tau\eta_b^{l,t} \mathbf{x}_b^{l,t}))}{\tau\eta_b^{l,t}}\in \partial\Psi_l(\boldsymbol{\gamma}_b^{t+1}).
\end{align}
Hence, we can arrive at the following subgradient estimator $\mathbf{x}_b^{l,t+1}$ such that \eqref{mainstep} holds:
\begin{equation}\label{xbl}
  \mathbf{x}_b^{l,t+1}=\mathbf{x}_b^{l,t}+\frac{1}{\tau\eta_b^{l,t}}(\mathbf{z}_b^t-\boldsymbol{\gamma}_b^{t+1}),
\end{equation}
where the right hand side (RHS) of \eqref{xbl} is obtained by replacing the proximal step in \eqref{xwl} by $\boldsymbol{\gamma}_b^{t+1}$ and further reorganizing the left hand side (LHS) of formula \eqref{xwl}. Consequently, the subgradient estimator $\mathbf{x}_b^t$ in \eqref{zt} can be updated as
\begin{equation}\label{full}
  \mathbf{x}_b^{t+1}=\mathbf{x}_b^{t}+c_{lb}^t(\mathbf{x}_b^{l,t+1}-\mathbf{x}_b^{l,t}),
\end{equation}
which exploits the fact that $ \mathbf{x}_b^{t}=\sum_{l=1}^{|\mathcal{N}_b|_c}c_{lb}^t\mathbf{x}_b^{l,t}$ and as stated in \eqref{mainstep}, only a single selected $\mathbf{x}_b^{l,t}$ is updated in each iteration $t$.

\subsection{Derivation of Algorithm Recursions}
Since the proximal operator needs to be calculated at each iteration in \eqref{zt} and \eqref{mainstep}, it is important to derive closed-form expressions for evaluating $\boldsymbol{\gamma}_b^{t+1}$ and $\mathbf{z}_b^t$ exactly. We start by calculating the gradient of local estimator $f(\boldsymbol{\gamma}_b)$ in \eqref{eqr}.
According to the well-known Sherman-Morrison rank-1 update identity \cite{sher}, we obtain
\begin{align}\label{sdd}
\left ( \boldsymbol{\Sigma}_{bn}+{\gamma}_{bn}\mathbf{s}_n\mathbf{s}_n^H\right )^{-1}=\boldsymbol{\Sigma}_{bn}^{-1}-
\frac{{\gamma}_{bn}\boldsymbol{\Sigma}_{bn}^{-1}\mathbf{s}_n\mathbf{s}_n^H\boldsymbol{\Sigma}_{bn}^{-1}}{1+{\gamma}_{bn}\mathbf{s}_n^H\boldsymbol{\Sigma}_{bn}^{-1}\mathbf{s}_n},
\end{align}
with
\begin{equation}\label{sib}
  \boldsymbol{\Sigma}_{bn}=\boldsymbol{\Sigma}_b- {\gamma}_{bn}\mathbf{s}_n\mathbf{s}_n^H,
\end{equation}
where ${\gamma}_{bn}$ is the $n$th element of $\boldsymbol{\gamma}_{b}$. Applying the well-known determinant identity{\color{blue}\footnote{Determinant identity \cite{deter} states that if $\mathbf{A}$ and $\mathbf{B}$ are matrices of sizes $m\times n$ and $n\times m$, then $\text{det}(\mathbf{I}+\mathbf{A}\mathbf{B})=\text{det}(\mathbf{I}+\mathbf{B}\mathbf{A})$.}} to $\boldsymbol{\Sigma}_{bn}+{\gamma}_{bn}\mathbf{s}_n\mathbf{s}_n^H$ yields
\begin{eqnarray}\label{sdd1}
\text{det}(\boldsymbol{\Sigma}_{bn}+{\gamma}_{bn}\mathbf{s}_n\mathbf{s}_n^H)=(1+{\gamma}_{bn}\mathbf{s}_n^H\boldsymbol{\Sigma}_{bn}^{-1}\mathbf{s}_n)\text{det}(\boldsymbol{\Sigma}_{bn}).
\end{eqnarray}
Then, substituting \eqref{sdd} and \eqref{sdd1} into \eqref{eqr} and taking the derivative of $f(\boldsymbol{\gamma}_b)$ with respect to ${\gamma}_{bn}$ leads to
\begin{eqnarray}\label{gsi}
\bigtriangledown f(\gamma_{bn})=\frac{\mathbf{s}_n^H\boldsymbol{\Sigma}_{bn}^{-1}\mathbf{s}_n}{1+{\gamma}_{bn}\mathbf{s}_n^H\boldsymbol{\Sigma}_{bn}^{-1}\mathbf{s}_n}-\frac{\mathbf{s}_n^H\boldsymbol{\Sigma}_{bn}^{-1}\hat{\boldsymbol{\Sigma}}_{b\mathbf{y}}\boldsymbol{\Sigma}_{bn}^{-1}\mathbf{s}_n}{(1+{\gamma}_{bn}\mathbf{s}_n^H\boldsymbol{\Sigma}_{bn}^{-1}\mathbf{s}_n)^2}.
\end{eqnarray}
Correspondingly, the gradient $\bigtriangledown f(\boldsymbol{\gamma}_b^t)$ can be derived as $\bigtriangledown f(\boldsymbol{\gamma}_{b}^t)=\text{col}\{\bigtriangledown f(\gamma_{b1}^t), \cdots,\bigtriangledown f(\gamma_{bN}^t)\}$.
Observe that the full gradient of $f(\boldsymbol{\gamma}_{b})$ over $\boldsymbol{\gamma}_{b}$ is calculated in each iteration, resulting in high computational complexity.
To perform as few gradient calculations as possible for striking a trade-off between algorithm exploration and accuracy, this paper adopts a non-dense gradient update rule where the full gradient is computed with a certain probability $0 <\bar{p} \leq 1$ resulting in a cheap per-iteration cost. Specifically, we construct a function of the current update in the $t$ iteration for computing the gradient as follow
\begin{eqnarray}\label{gt}
\widetilde{\bigtriangledown f}(\boldsymbol{\gamma}_{b}^t) = \left\{ \begin{array}{l}
 \bigtriangledown f(\boldsymbol{\gamma}_{b}^t),~\text{with probability} ~\bar{p},\\
\mathbf{0},~~~~\text{with probability} ~1-\bar{p},
\end{array} \right.
\end{eqnarray}
Indeed, since each AP updates the gradient with probability $\bar{p}$ at each iteration, all these updates add up to a dense one.

By substituting \eqref{gt} into \eqref{zt} and calculating the proximal step of $g(\boldsymbol{\gamma}_b)$, we can obtain the following intermediate recursion
\begin{eqnarray}\label{zbt}
\!\!\!\!\!\!\!\!\!\!\!\!&&\!\!\!\!\!\!\!\!\!\!\!\! \mathbf{z}_b^{t}= \boldsymbol{\varsigma}_{b}^t-\eta_b^t\beta \text{col}\left\{ \frac{ \varsigma_{b1}^t}{ \left \| \mathbf{R}_b^t(1,:) \right \|_2 },\cdots,\frac{ \varsigma_{bN}^t}{ \left \| \mathbf{R}_b^t(N,:) \right \|_2 }\right\},
\end{eqnarray}
with $
  \boldsymbol{\varsigma_{b}}^t= \boldsymbol{\gamma}_b^t-\eta_b^t\widetilde{\bigtriangledown f}(\boldsymbol{\gamma}_{b}^t)-\tau\eta_b^t\mathbf{x}_b^t$,
where $\varsigma_{bn}^t$ is the $n$th element of $\boldsymbol{\varsigma}_{b}^t$.

Note that the step size $\eta_b^t$ in \eqref{zbt} is an essential hyper-parameter, which should be designed for improving the detection performance. In particular, a large step size leads to a faster convergence initially but early saturation,
while small step size slows the convergence, but postpone the saturation. In fact, it is a difficult task in practice as one approximate step size is determined by the Lipschitz constant which is hard to estimate. The authors in \cite{imp} proposed the adaptive step size for stochastic gradient based on the quasi-Newton property:
\begin{eqnarray}\label{seo}
\eta_b^t=\frac{\left \| \boldsymbol{\gamma}_{b}^t-\boldsymbol{\gamma}_{b}^{t-1} \right \|_2^2}{ (\boldsymbol{\gamma}_{b}^t-\boldsymbol{\gamma}_{b}^{t-1})^H(\widetilde{\bigtriangledown f}(\boldsymbol{\gamma}_{b}^t)-\widetilde{\bigtriangledown f}(\boldsymbol{\gamma}_{b}^{t-1})) },
\end{eqnarray}
which is the local estimation of Lipschitz constant and approximates the inverse of Hessian matrix of $f(\cdot)$ at $\boldsymbol{\gamma}_{b}^t$. However, the denominator of \eqref{seo} is potentially small due to the fact that $\boldsymbol{\gamma}_{b}^t$ is highly sparse. Also, the gradient even remains constant between two consecutive iteration indices due to the application of gradient in a probabilistic manner in \eqref{gt}. In consequence,  $\tilde{\eta}_b^t$ may approach infinity. To avoid this case, we introduce the term $\epsilon\left \| \boldsymbol{\gamma}_{b}^t-\boldsymbol{\gamma}_{b}^{t-1}\right\|$ to the denominator of \eqref{seo} to
keep the its denominator positive and control the lower bound of the denominator in each iteration. The resulting step size is explicitly given by
\begin{align}\label{sep}
\eta_b^t=\frac{\left \| \boldsymbol{\gamma}_{b}^t-\boldsymbol{\gamma}_{b}^{t-1} \right \|_2^2}{\left| (\boldsymbol{\gamma}_{b}^t\!-\!\boldsymbol{\gamma}_{b}^{t-1})^H(\widetilde{\bigtriangledown f}(\boldsymbol{\gamma}_{b}^t)\!-\!\widetilde{\bigtriangledown f}(\boldsymbol{\gamma}_{b}^{t-1})) \right|\!+\! \epsilon\left \| \boldsymbol{\gamma}_{b}^t\!-\!\boldsymbol{\gamma}_{b}^{t-1} \right \|_2^2},\nonumber\\
\end{align}
where $\epsilon\geq 0$ is the adjustment parameter. Noting that \eqref{sep} can adaptive tune the step size without imposing much extra computational complexity burden.

Now, we turn to derive the recursion of $\boldsymbol{\gamma}_b^{t}$ in \eqref{mainstep}. Since $\Psi_l(\cdot)$ in \eqref{mainstep} is fully separable, its proximal operator can be evaluated component-wise, which is soft thresholding given by
\begin{align}\label{pos1}
\gamma_{bn}^{t+1}=\left\{\begin{array}{l}
\min\left(\tau \eta_b^{l,t}\frac{z_{bn}^t+\tau\eta_b^{l,t} x_{bn}^{l,t}-{\gamma}_{ln}^t}{|z_{bn}^t\!+\!\tau\eta_b^{l,t} x_{bn}^{l,t}-{\gamma}_{ln}^t|},z_{bn}^t+\tau\eta_b^{l,t} x_{bn}^{l,t}\right)\\
+z_{bn}^t+\tau\eta_b^{l,t} x_{bn}^{l,t},~\text{if}~~ z_{bn}^t+\tau\eta_b^{l,t} x_{bn}^{l,t}\neq0,\\
0,~~~~~~~~~~~~~~~~~~~\text{if}~~ z_{bn}^t+\tau\eta_b^{l,t} x_{bn}^{l,t}=0,
\end{array} \right.
\end{align}
where the minimum selection operator is to preserve the positivity of $\gamma_{bn}$. Here, $z_{bn}^t$ and $x_{bn}^{l,t}$ are the $n$th entry of vectors $\mathbf{z}_b^t$ and $\mathbf{x}_b^{l,t}$, respectively. After getting $\gamma_{bn}^{t+1}$, the covariance matrix $\boldsymbol{\Sigma}_{b}$ in \eqref{sib} can be updated according to the distance between $\gamma_{bn}^{t+1}$ and $\gamma_{bn}^{t}$, which is given by
\begin{equation}\label{sm}
  \boldsymbol{\Sigma}_b^{t+1}=\boldsymbol{\Sigma}_b^{t}+(\gamma_{bn}^{t+1}-\gamma_{bn}^{t})\mathbf{s}_n\mathbf{s}_n^H.
\end{equation}
From \eqref{full} and \eqref{pos1}, the estimation performance depends, to a great extent, on the cooperation strategy specified by the combiner $c_{lb}^t$. Similar to work \cite{tls}, this paper adopts the following adaptive combiner
\begin{eqnarray}\label{clb}
c_{lb}^t = \left\{\begin{array}{l}
\frac{2}{\left | \mathcal{N}_b^{-} \right |_c}\frac{1}{1+\exp(\rho \|\boldsymbol{\gamma}_b^{t-1}-\boldsymbol{\gamma}_l^{t-1}\|_2 )},~l\in \mathcal{N}_b^{-},\\
1-\sum _{l\in \mathcal{N}_b^{-}}c_{lb}^t,~~~~~~~~~~~~~~~l=b,\\
0, ~~~~~~~~~~~~~~~~~~~~~~~~~~~~~~~l\notin  \mathcal{N}_b,
\end{array} \right.
\end{eqnarray}
where $\rho>0$ is a large constant set beforehand. Note that the term $\|\boldsymbol{\gamma}_b^{t-1}-\boldsymbol{\gamma}_l^{t-1} \|_2$ in \eqref{clb} accounts for the distance between the local estimates of the $b$th AP and its $l$th neighbor. The combiner $c_{lb}^t$ is inversely proportional to such a distance. When the distance defined above between two APs is large, the $b$th AP tends to decrease the value of combiner, or even discard the information from this neighbor. Conversely, the $b$th AP would increase the combiner when the distance of estimation between two APs is small.

Once the estimate $\{\hat{\boldsymbol{\gamma}}_b\}_{b=1}^B$ of the device state vectors is obtained, we employ the element-wise thresholding at each AP to determine $\chi_n$ from $\hat{\gamma}_{bn}$, i.e., $\chi_n=1$, if $\hat{\gamma}_{bn}>\imath_b$, and $\chi_n=0$ otherwise, where the threshold $\imath_b=\omega_b\sigma^2$ for a tunable parameter $\omega_b>0$. For clarity, the pseudo-code of the cooperative activity detection (CAD) algorithm for sourced random access is summarized in Algorithm 1. Different from the cooperative MIMO \cite{multicell}, where each BS performs signal processing locally, the vectors of log-likelihood ratio are then forwarded to the central unit for decision making, the proposed CAD algorithm exchanges device state vector iteratively among neighboring APs. In this context, the access latency in cooperative MIMO is longer than that of the proposed CAD algorithm. On the other hand, the proposed CAD algorithm is more reliable and robust to AP and/or fronthaul link failure compared with cooperative MIMO.

\begin{algorithm}[h]
\caption{CAD for Sourced Random Access}
\label{alg1}
\begin{algorithmic}[1]
\STATE \textbf{Input}: $\{\mathbf{Y}_b\}_{b=1}^B$, $\mathbf{S}$, $\{\hat{\boldsymbol{\Sigma}}_{b\mathbf{y}}=\frac{1}{M}\mathbf{Y}_b\mathbf{Y}_b^H\}_{b=1}^B$, step size $\{\eta_b^0\}_{b=1}^B$, and total iterations $T$
\STATE \textbf{Initialization}: $\{\boldsymbol{\gamma}_b^{0}=\mathbf{0}\}_{b=1}^B$, $\{\boldsymbol{\Sigma}_b^0=\sigma^2\mathbf{I}\}_{b=1}^B$, $\{\mathbf{x}_b^{l,0}, l\in \mathcal{N}_b\}_{b=1}^B$, $\{c_{lb}^0\}_{b=1}^B$, $\{\mathbf{x}_b^0= \sum_{l\in \mathcal{N}_b}c_{lb}^0\mathbf{x}_b^{l,0}\}_{b=1}^B$ \\
\FOR{$t=1 : T$}
\STATE {\bf{for each AP $b$:}}
\STATE {\bf{Adaptation:}}
\STATE Compute $\widetilde{\bigtriangledown f}(\boldsymbol{\gamma}_{b}^t)$ based on \eqref{gt}
\STATE Compute $\mathbf{z}_b^{t}
$ based on \eqref{zbt}
\STATE Compute adaptive step size $\eta_b^t$ based on \eqref{sep}
\STATE Choose $l$ randomly from the set $\mathcal{N}_b$ with probabilities $\{p_1,p_2,\cdots,p_{\left | \mathcal{N}_b \right |_c}\}$
\STATE Compute adaptive combiner $c_{lb}^t$ based on \eqref{clb}
\STATE Compute $\eta_b^{l,t}=\frac{c_{lb}^t\eta_b^t}{ p_l}$
\STATE Randomly select a permutation $t_1,t_2,\cdots,t_N$ of the
coordinate indices $\{1,2,\cdots,N\}$ of $\boldsymbol{\gamma}_{b}^t$
\FOR{$n=1 : N$}
\STATE  Update $\gamma_{bt_n}^{t+1}$ based on \eqref{pos1}
\STATE $\boldsymbol{\Sigma}_b^{t+1}=\boldsymbol{\Sigma}_b^{t}+(\gamma_{bt_n}^{t+1}-\gamma_{bt_n}^{t})\mathbf{s}_{t_n}\mathbf{s}_{t_n}^H$
\ENDFOR
\STATE Compute $\mathbf{x}_b^{l,t+1}$ based on \eqref{xbl}
\STATE Compute $\mathbf{x}_b^{t+1}$ based on \eqref{full}
\STATE {\bf{Communication:}}
\STATE Transmit $\boldsymbol{\gamma}_{b}^t$ to its one-hop neighbor AP
\ENDFOR
\STATE \textbf{Output}: $\{\boldsymbol{\gamma}_b^{t+1}\}_{b=1}^B$
\end{algorithmic}
\end{algorithm}

\section{Cooperative Unsourced Random Access}
In this section, we design an unsourced random access scheme for 6G cell-free wireless networks. Indeed, most works on unsourced random access comprise an inner code for recovering the submessage and an outer code for stitching the submessages \cite{ura-single-ante, unsourced}. In this context, the proposed CAD algorithm in Section III can be employed as an inner decoder of inner code for unsourced random access. Hence, we can provide a unified algorithm for both sourced and unsourced random access paradigms.

For the proposed cooperative unsourced random access scheme, the message of each device is split into several subblocks, which are independently transmitted and cooperatively recovered at all APs, and then to be integrated to reconstruct the original message. Herein, the transmission and the recovery of the subblock messages are called cooperative inner code, while the split and integration of the message is called outer code. In what follows, we provide the details about the cooperative inner and outer code by exploiting the characteristics of 6G cell-free wireless networks.

\subsection{Cooperative Inner Code}
It is assumed that each active device has a message of length $\bar{b}=uR$ bits to send with $u=\mathcal{Z}L$, where $u$ denotes the length of signal samples, $\mathcal{Z}$ denotes the number of subblocks, $R$ denotes the sample rate, and $L$ denotes the subslot length. The active devices synchronously transmit their messages over the coherence time block to all the APs in 6G cell free wireless network. Then, each AP $b$ produces a list $\mathcal{L}_b$ of the transmitted messages $\{m_{b,k}:k\in \mathcal{K}\}$ of the active devices. For some integer $J > 0$, the $\bar{b}$-bit message is divided into $\mathcal{Z}$ subblocks of size $q_1,q_2,\cdots,q_{\mathcal{Z}}$ satisfying the following conditions: $\sum_{i}q_i=\bar{b}$, $q_1=J$, and $q_i<J$ for all $i=2, \cdots, \mathcal{Z}$. Herein, all subblocks $i=2,\cdots,\mathcal{Z}$ are augmented to size $J$ by appending the parity bits and the pseudo-random parity-check equations are identical for all devices. Such a systematic linear block code based on random parity checks is known as the tree code proposed in \cite{ura-single-ante}.

Then, the inner code is used to transmit in sequence the $\mathcal{Z}$ subblocks under the subslot length $L$. Let $\mathbf{S}\in \mathbb{C}^{L\times 2^J}=[\mathbf{s}_1,\mathbf{s}_2,\cdots, \mathbf{s}_{2^J}]$ represent a common codebook for all devices with columns being the normalized codewords such that $\|\mathbf{s}_e\|_2^2=L, \forall e \in \{1,2,\cdots,2^J\}$. For subblock $i$, let $e_k(i)$ denote the $J$-bit messages produced by the $k$th active device characterized by integers in $\{1,2,\cdots, 2^J \}$. The active device $k$ simply sends the column $\mathbf{s}_{e_k(i)}$ of the coding matrix $\mathbf{S}$ to all APs. As such, the received signal at the $b$th AP during the $i$th subblock can be expressed as
\begin{align}\label{us}
\mathbf{Y}_{b,i}&=\sum_{k\in\mathcal{K}}\sqrt{g_{b,k}}\mathbf{s}_{e_k(i)}\mathbf{h}_{b,k}^T+\mathbf{W}_{b,i}\nonumber \\
&=\mathbf{S}\boldsymbol{\Xi}_{b,i}\mathbf{G}_b^{1/2}\mathbf{H}_{b}+\mathbf{W}_{b,i}, ~\forall i \in \{1,2,\cdots,\mathcal{Z}\},
\end{align}
where $g_{b,k}$, $\mathbf{h}_{b,k} \in \mathbb{C}^M$, and $\mathbf{H}_{b} \in \mathbb{C}^{N \times M}$ are the large-scale fading component, the small-scale fading vector, and the corresponding small-scale fading matrix (Gaussian i.i.d. entries $\thicksim\mathcal{CN}(0,1)$) as described in Section III, $\mathbf{G}_b=\text{diag}(g_{b,1},\cdots,g_{b,N})$ denotes the diagonal large-scale fading matrix, $\mathbf{W}_{b,i}$ is the AWGN matrix with i.i.d. entries $\thicksim\mathcal{CN}(0,\sigma^2)$, and $\boldsymbol{\Xi}_{b,i}\in \{0,1\}^{2^J \times N}$ is a binary selection matrix where for the active device $k$ the corresponding column is all-zero but has a single one in position $e_k(i)$, while for all inactive devices the corresponding columns contain all zeros.

The elements of the $r$th row of the matrix $\boldsymbol{\Xi}_{b,i}\mathbf{G}_b^{1/2}\mathbf{H}_{b}$ follow the distribution of $\mathcal{CN}(0,\sum_{k\in \mathcal{K}}g_{b,k}\xi_{r,k}^{b,i})$, where $\xi_{r,k}^{b,i}$ denotes the $(r,k)$-th element of $\boldsymbol{\Xi}_{b,i}$, which equals to one if $r=e_k(i)$ and zero otherwise. Let $\gamma_{bi}^r=\sum_{k\in \mathcal{K}}g_{b,k}\xi_{r,k}^{b,i}$ and ${\boldsymbol{\Gamma}}_{b,i}=\text{diag}(\gamma_{bi}^1,\cdots,\gamma_{bi}^{2^J})\in \mathbb{C}^{2^J \times 2^J}$, then, the received signal can be transformed as
\begin{eqnarray}\label{us1}
\mathbf{Y}_{b,i}&=&\mathbf{S}\boldsymbol{\Gamma}_{b,i}^{\frac{1}{2}}\bar{\mathbf{H}}_b+\mathbf{W}_{b,i}, i=1,2,\cdots,\mathcal{Z},
\end{eqnarray}
where $\bar{\mathbf{H}}_b$ has the i.i.d. elements $\thicksim\mathcal{CN}(0,1)$. Note that for a given subblock $i$, \eqref{us1} has the same form as \eqref{aps}. Moreover, it is seen that the dimensions of the variable matrices in \eqref{us1} are independent of the number of total devices $N$.

The decoding of the inner code is equivalent to recovering the non-zero elements of
the diagonal matrix $\boldsymbol{\Gamma}_{b,i}$ from the noisy measurements $\mathbf{Y}_{b,i}$ for
all APs and subblocks. Under this context, given subblock $i$, the codeword state vectors $\{\boldsymbol{\gamma}_{bi}=[\gamma_{bi}^1,\cdots,\gamma_{bi}^{2^J}]\}_{b=1}^{B}$ has the same characteristics as the device state vectors $\{\boldsymbol{\gamma}_{b}\}_{b=1}^B$. Specifically, they are jointly sparse and the amplitudes of the nonzero entries at the same position of neighboring APs are similar to each other. It is interesting to find that the decoding of inner code is the same as the cooperative activity detection for sourced random access. Hence, the proposed CAD algorithm in the last section can be employed as the cooperative decoding algorithm of the inner code.

\subsection{Outer Code in Cooperative Systems}
Once the estimation of the codeword state vector, $\hat{\boldsymbol{\gamma}}_{b}^i=[\hat{\gamma}_{b1}^i,\hat{\gamma}_{b2}^i,\cdots,\hat{\gamma}_{b2^J}^{i}]^T$, is obtained, it is then passed to the outer decoder for further identify all possible subblock messages by separating data and parity bits. In this paper, we utilize the concatenated code as the outer code. In what following, we provide a decoding approach for the outer code.

For the $b$th AP, the set of active columns of codebook $\mathbf{S}$ at subblock $i$ is given by
\begin{eqnarray}\label{sub}
\mathcal{C}_{b,i}=\{r\in\{1,2,\cdots,2^J\}:\hat{\gamma}_{br}^i\geq\imath_{b,i}\},
\end{eqnarray}
where the threshold is defined as $\imath_{b,i}=\nu_{b,i}\sigma^2$ with a tuneable parameter $\nu_{b,i}>0$ for AP $b$ and subblock $i$. The set $\mathcal{C}_{b,i}$ forms the list of submessages at the $b$th AP and subblock $i$. For the $b$th AP, since the subblocks contain parity bits, not all message sequences in $\mathcal{C}_{b,1}\times \mathcal{C}_{b,2}\times\cdots\times\mathcal{C}_{b,\mathcal{Z}}$ are needed.
Therefore, the outer decoder is used to identify all possible message sequences. In specific, starting from $i=1$ and proceeding in order, the integer indices $\mathcal{C}_{b,i}$ is converted back to their binary representation, and then the sequences satisfying the parity are output as the original messages. A description of the CAD for unsourced random access is summarized in Algorithm 2.
\begin{algorithm}[h]
\caption{CAD for Unsourced Random Access}
\label{alg1}
\begin{algorithmic}[1]
\STATE \textbf{Input}: $\{\mathbf{Y}_b\}_{b=1}^B$ and $\mathbf{S}\in \mathbb{C}^{L\times 2^J}$
\STATE {\bf{Outer encoder}}: The $\mathcal{Z}$ subblocks of messages are encoded via a tree code
\STATE {\bf{Inner encoder}}: Transmit in sequence the $\mathcal{Z}$ subblocks to all APs based on $\mathbf{S}$
\FOR{$i=1 : \mathcal{Z}$}
\STATE {\bf{for each AP $b$:}}
\STATE {\bf{Cooperative Inner decoder}}: Inner decoding with the Algorithm 1 to obtain the estimation $\hat{\boldsymbol{\gamma}}_{b}^i$
\ENDFOR
\STATE {\bf{Outer decoder}}: Find the set $\{\mathcal{C}_{b,i}\}_{b=1}^B$ by hard threshold and stitch together the sequence of submessages
\STATE \textbf{Output}: The list $\{\mathcal{L}_b\}_{b=1}^B$ of the transmitted messages
\end{algorithmic}
\end{algorithm}

\section{Performance Analysis}
As mentioned earlier, the CAD algorithm is the key of both cooperative sourced and unsourced random access. In this section, we provide the analysis of computational complexity, communication cost, and the convergence of the proposed CAD algorithm.
\subsection{Computational Complexity and Communication Cost}
In what follows, the computational complexity and communication cost of the proposed CAD algorithm is analyzed. In each iteration of Algorithm 1, for an arbitrary AP, the computational complexity mainly arises from the matrix multiplication, and the overall computational complexity of the CAD algorithm is $\mathcal{O}(L^2)$. Although the computational complexity of sample covariance $\hat{\boldsymbol{\Sigma}}_{b\mathbf{y}}$ is $\mathcal{O}(L^2M)$, it only needs to be calculated once at each time slot before the start of iteration. For an arbitrary AP, we compare the proposed algorithm with two detection algorithms from the perspective of the computational complexity, including the AMP algorithm \cite{multicell} and the OAMP-MMV algorithm \cite{oamp}. It can be seen in Table I that the computational complexity of the proposed CAD algorithm is superior to other two algorithms in mMTC applications.

\newcommand{\tabincell}[2]{\begin{tabular}
{@{}#1@{}}#2\end{tabular}}
\begin{table}[h]
\centering
\caption{Computation Complexity of the Considered Schemes.}
\label{tab1}
\begin{tabular}{lcccc}
\hline
\tabincell{c}{Schemes}     &\tabincell{c}{Computational\\ complexity}
\\
\hline
Proposed CAD & $\mathcal{O}(TL^2+L^2M)$
  \\
AMP  & $\mathcal{O}(T(MNL+LM^2+NM^3))$\\
OAMP-MMV  & $\mathcal{O}(T(MN\log_2N+NM^3))$\\
\hline
\end{tabular}
\end{table}

For the communication cost, in each iteration, each AP needs to transmit $N$-dimensional intermediate $\boldsymbol{\gamma}_{b}^t$. Thus, for all APs, the CAD algorithm needs to exchange $N\sum_{b=1}^B\left|\mathcal{N}_b^{-}\right|_c$ number of parameters. For the activity detection in cooperative unsourced random access, the role of the total number of potential devices $N$ in the device activity detection problem is replaced by the number of messages $2^J\ll N$ with the small size $J$. Thus the communication cost of the CAD algorithm reduce from $N\sum_{b=1}^B\left|\mathcal{N}_b^{-}\right|_c$ to $2^J\sum_{b=1}^B\left|\mathcal{N}_b^{-}\right|_c$.

\emph{Remark 1}: It is interesting to emphasize that the computational complexity and the communication cost at each iteration of the CAD algorithm do not grow as the number of antennas at each AP, $M$, increases. Moreover, cooperative unsourced random access implies that both the computational complexity and the communication cost of the proposed CAD algorithm do not grow by increasing the total number of potential devices. Instead, the overhead is determined by the length of subblock $J$. In practical applications, the wireless fronthaul links among APs usually have limited capacity in 6G cell-free wireless network. Thus, $J$ can be flexibly designed to strike a balance between the detection accuracy and required signaling overhead. In addition, a small number of APs can be set for cooperation, as will be verified by simulations in Section VI, where the activity detection error can approach zero with the existence of limited fronthaul links.

\subsection{Convergence Analysis}
In this subsection, we establish the convergence (to a stationary point of $\boldsymbol{\gamma}_{b}^t$) of the proposed CAD algorithm. Before giving a more detailed convergence result, let us first provide the following lemma which is instrumental in analyzing the convergence of the proposed algorithm.

\emph{Lemma 1}: Let $\boldsymbol{\zeta}_b^t= \boldsymbol{\gamma}_b^t-\bar{p}\eta_b^t\triangledown f(\boldsymbol{\gamma}_b^t)$ and $\boldsymbol{\zeta}_b^*= \boldsymbol{\gamma}_b^*-\bar{p}\eta_b^t\bigtriangledown f(\boldsymbol{\gamma}_b^*)$, where  $\boldsymbol{\gamma}_b^*$ is the optimal solution of minimizing the problem in \eqref{ob1}. When $f(\cdot)$ is $\mathcal{L}_f$-smooth (i.e., $\|\bigtriangledown f(\boldsymbol{\gamma}_{b}^t)-\bigtriangledown f(\boldsymbol{\gamma}_{b}^{t-1}\|\leq \mathcal{L}_f\|\boldsymbol{\gamma}_{b}^t-\boldsymbol{\gamma}_{b}^{t-1}\|$) defined in \cite{boy}, we have
\begin{eqnarray}\label{lem1}
\mathbb{E}\left \{ \left \| \boldsymbol{\zeta}_b^t-\boldsymbol{\zeta}_b^* \right \|_2^2\right\}\leq \mathbb{E}\left \{\left  \| \boldsymbol{\gamma}_b^t-\boldsymbol{\gamma}_b^* \right \|_2^2\right\}-\bar{p}\varpi \eta_{b}^td_f(\boldsymbol{\gamma}_b^t,\boldsymbol{\gamma}_b^*), \end{eqnarray}
for an arbitrary AP $b$, the smoothness constant $\mathcal{L}_f$, and some constant $\varpi >0$,  if the step sizes are
chosen to satisfy
\begin{eqnarray}\label{ss}
\frac{1}{\mathcal{L}+\epsilon}\leq \eta_{b}^t<\min\left(\frac{2}{\mathcal{L}_f},\frac{1}{\epsilon}\right), b=1,2,\cdots,B,
\end{eqnarray}
where $d_f(\boldsymbol{\gamma}_b^t,\boldsymbol{\gamma}_b^*)$ denotes the Bregman divergence of a function $f(\cdot)$ and is given by
$d_f(\boldsymbol{\gamma}_b^t,\boldsymbol{\gamma}_b^*):=f(\boldsymbol{\gamma}_b^t)-f(\boldsymbol{\gamma}_b^*)-
\left \langle\triangledown f(\boldsymbol{\gamma}_b^t) ,\boldsymbol{\gamma}_b^t-\boldsymbol{\gamma}_b^* \right \rangle$.

\begin{IEEEproof}
Please refer to Appendix A.
\end{IEEEproof}

As stated, the bound \eqref{lem1} in Lemma 1, which depends on the Bregman divergence, will play an important role in verifying the convergence of the proposed CAD algorithm.
We then turn to prove the existence and the optimality of the fixed points of step 7 and step 13 in Algorithm 1.

\emph{Lemma 2}: For an arbitrary AP $b$, the solution $\boldsymbol{\gamma}_b^*$ of minimizing the problem in \eqref{ob1} exists for recursions step 7 and step 13 in Algorithm 1, i.e., it holds that
\begin{align}
 &\boldsymbol{\gamma}_b^*=\text{prox}_{\beta\eta_b^t g}(\boldsymbol{\gamma}_b^*-\eta_b^t\widetilde{\bigtriangledown f}(\boldsymbol{\gamma}_b^*)-\tau\eta_b^t\mathbf{x}_b^*),\label{lem2}\\    &\boldsymbol{\gamma}_b^{*}=\text{prox}_{\tau\eta_b^{l,t} \Psi_l}(\mathbf{z}_b^*+\tau\eta_b^{l,t} \mathbf{x}_b^{l,*}),\label{lem3}
\end{align}
for any $\eta_b^t$ and $\eta_b^{l,t}$, if
\begin{eqnarray}\label{hold}
  &&\bigtriangledown f(\boldsymbol{\gamma}_{b}^*)+\tau\sum_{l\in \mathcal{N}_b}c_{lb}^t\mathbf{x}_b^{l,*}+\beta\boldsymbol{\varrho }_b^{*}=\mathbf{0},
\end{eqnarray}
holds for the vectors $\mathbf{x}_b^* \in \partial \Psi (\boldsymbol{\gamma}_b^*)$, $\mathbf{x}_b^{l,*}\in \partial \Psi_l (\boldsymbol{\gamma}_b^*)$, and $\boldsymbol{\varrho }_b^{*}\in \partial g(\boldsymbol{\gamma}_b^*)$. Condition \eqref{hold} means that when the optimality is obtained, the subgradient of local cost function in \eqref{ob1} reaches $0$.

\begin{IEEEproof}
Please refer to Appendix B.
\end{IEEEproof}

Note that there exists a particular fixed point $\boldsymbol{\gamma}_b^{*}$ and we will show that the iterations of $\boldsymbol{\gamma}_b^{t}$
converge to this particular fixed point in the following.
We collect the information from across all APs into block
vectors and matrices. In particular, for the $b$th AP, we define the mean of the variables $\boldsymbol{\gamma}_b^t$ with respect to time series $t$ as
$
\bar{\boldsymbol{\gamma}}_b^t=\frac{1}{t}\sum_{j=0}^{t-1}\boldsymbol{\gamma}_b^t$.
The global stochastic quantities are defined as
\begin{eqnarray}\label{global}
&&\boldsymbol{\gamma}^*=\text{col}\{\boldsymbol{\gamma}_1^*,\boldsymbol{\gamma}_2^*,\cdots,\boldsymbol{\gamma}_{B}^*\},\nonumber\\ &&\bar{\boldsymbol{\gamma}}^t=\text{col}\{\bar{\boldsymbol{\gamma}}_1^t,\bar{\boldsymbol{\gamma}}_2^t,\cdots,\bar{\boldsymbol{\gamma}}_{B}^t\},\nonumber\\
&&\mathbf{h}^t=\text{col}\{d_f(\bar{\boldsymbol{\gamma}}_1^t,\boldsymbol{\gamma}_1^*),d_f(\bar{\boldsymbol{\gamma}}_2^t,\boldsymbol{\gamma}_2^*),\cdots,d_f(\bar{\boldsymbol{\gamma}}_{B}^t,\boldsymbol{\gamma}_{B}^*)\}.~~~~~
\end{eqnarray}
The next result provides a convergence rate of the proposed CAD algorithm.

\emph{Theorem 1:} If the step size satisfies $\frac{1}{\mathcal{L}_f+\epsilon}\leq \eta_{b}^t<\min(\frac{2}{\mathcal{L}_f},\frac{1}{\epsilon})$, $b=1,2,\cdots,B$, the inequality
\begin{eqnarray}\label{tem1}
\mathbb{E}\left \{ \mathbf{h}^t\right\}\leq \frac{1}{\bar{p}\varpi t}\frac{1}{\boldsymbol{\eta}^t}\odot \boldsymbol{\mathcal{W}}^0,
\end{eqnarray}
holds where $\frac{1}{\boldsymbol{\eta}^t}=\text{col}\{\frac{1}{ \eta_1^t},\frac{1}{\eta_2^t},\cdots,\frac{1}{\eta_B^t}\}$, the expectation is taken over random choice of neighbor AP $l$ of the $b$th AP, and $\boldsymbol{\mathcal{W}}^0=\text{col}\{\mathcal{W}_1^0,\mathcal{W}_2^0,\cdots,\mathcal{W}_{N}^0\}$ with
\begin{eqnarray}
\mathcal{W}_b^0=\|\boldsymbol{\gamma}_b^0-\boldsymbol{\gamma}_b^*\|_2^2+\tau^2\sum_{l \in \mathcal{N}_b}(\eta_b^{l,0})^2\| \mathbf{x}_b^{l,0}- \mathbf{x}_b^{l,*}\|_2^2.\label{average}
\end{eqnarray}

\begin{IEEEproof}
Similar to the Lyapunov function adopted in \cite{acce1} and \cite{acce}, which can be used to prove the stability of an equilibrium, we define the Lyapunov function for each AP of the CAD algorithm at iteration $t$ as
\begin{eqnarray}\label{lya}
\mathcal{W}_b^{t+1}&=&\mathbb{E}\left \{ \|\boldsymbol{\gamma}_b^{t+1}-\boldsymbol{\gamma}_b^*\|_2^2\right\}\nonumber\\
&+&\sum_{l\in \mathcal{N}_b}(\tau\eta_b^{l,t+1})^2\mathbb{E}\left \{ \|\mathbf{x}_b^{l,t+1}-\mathbf{x}_b^{l,*}\|_2^2\right\},
\end{eqnarray}
which measures both distance between $\boldsymbol{\gamma}_b^{t}$ and $\boldsymbol{\gamma}_b^{*}$ and the sum of the distance between $\mathbf{x}_b^{l,t}$ and $\mathbf{x}_b^{l,*}$ for all the neighbors of the $b$th AP.

For establishing the upper bound on the first term of $\mathcal{W}_b^t$, we start by applying the non-expansiveness property of the proximal map \cite{forward}, which implies
\begin{eqnarray}\label{core}
 \!\!\!\!\!\!&&\!\!\!\!\!\!(1+\frac{1}{\tau\eta_b^{l,t}\mathcal{L}_{\Psi_l}})\|\mathbf{x}-\text{prox}_{\tau\eta_b^{l,t} \Psi_l}(\mathbf{x})-(\mathbf{y}-\text{prox}_{\tau\eta_b^{l,t } \Psi_l}(\mathbf{y}))\|_2^2\nonumber\\
 \!\!\!\!\!\!&&\!\!\!\!\!\!+\|\text{prox}_{\tau\eta_b^{l,t} \Psi_l}(\mathbf{x})-\text{prox}_{\tau\eta_b^{l,t } \Psi_l}(\mathbf{y}))\|_2^2\leq \|\mathbf{x}-\mathbf{y}\|_2^2,
\end{eqnarray}
when the function $\Psi_l$ is non-smooth with the smoothness constant
$\mathcal{L}_{\Psi_l}=+\infty $ for any $\mathbf{x}$ and $\mathbf{y}$ \cite{simple}.
Considering step 13 and step 16 in Algorithm 1, we
substitute $\mathbf{x}=\mathbf{z}_b^t+\tau\eta_b^{l,t} \mathbf{x}_b^{l,t}$ and $\mathbf{y}=\boldsymbol{\gamma}_b^*+\tau\eta_b^{l,t} \mathbf{x}_b^{l,*}$ into the inequality in \eqref{core} with the optimality condition of Lemma 2:
\begin{eqnarray}
\!\!\!\!\!\!\!\!\!\!\!\!&&\!\!\!\!\!\!\!\!\!\!\!\!\|\boldsymbol{\gamma}_b^{t+1}-\boldsymbol{\gamma}_b^*\|_2^2+(1+\frac{1}{\tau\eta_b^{l,t}\mathcal{L}_{\Psi_l}}) \nonumber \\
\!\!\!\!\!\!\!\!\!\!\!\!&&\!\!\!\!\!\!\!\!\!\!\!\! \cdot \|\mathbf{z}_b^t+\tau\eta_b^{l,t} \mathbf{x}_b^{l,t}-\boldsymbol{\gamma}_b^{t+1}-(\boldsymbol{\gamma}_b^*+\tau\eta_b^{l,t} \mathbf{x}_b^{l,*}-\boldsymbol{\gamma}_b^*)\|_2^2= \nonumber \\
\!\!\!\!\!\!\!\!\!\!\!\!&&\!\!\!\!\!\!\!\!\!\!\!\!\|\boldsymbol{\gamma}_b^{t+1}\!-\!\boldsymbol{\gamma}_b^*\|_2^2\!+\!(1\!+\!\frac{1}{\tau\eta_b^{l,t}\mathcal{L}_{\Psi_l}})(\tau\eta_b^{l,t})^2 \|\mathbf{x}_b^{l,t+1}\!\!-\!\!\mathbf{x}_b^{l,*}\|_2^2 \label{c1y} \\
\!\!\!\!\!\!\!\!\!\!\!\!&&\!\!\!\!\!\!\!\!\!\!\!\!\leq \|\mathbf{z}_b^t+\tau\eta_b^{l,t} \mathbf{x}_b^{l,t}-(\boldsymbol{\gamma}_b^*+\tau\eta_b^{l,t} \mathbf{x}_b^{l,*})\|_2^2.\label{c1}
\end{eqnarray}
Taking the expectation of the term in \eqref{c1} with respect to the random selected neighbor AP $l$ and putting the relation $\eta_b^{l,t}=\frac{c_{lb}^t\eta_b^t}{p_l}$ into it, we obtain
\begin{eqnarray}\label{c2}
\!\!\!\!\!\!\!&&\!\!\!\!\!\!\!\mathbb{E}\left \{\|\mathbf{z}_b^t\!+\!\tau\eta_b^{l,t} \mathbf{x}_b^{l,t}\!-\!(\boldsymbol{\gamma}_b^*+\tau\eta_b^{l,t} \mathbf{x}_b^{l,*})\|_2^2\right\}\!=\!\|\mathbf{z}_b^t-\boldsymbol{\gamma}_b^*\|_2^2+\nonumber\\
\!\!\!\!\!\!\!&&\!\!\!\!\!\!\! (\tau\eta_b^t)^2\sum_{l\in \mathcal{N}_b}\frac{(c_{lb}^t)^2}{p_l}\|\mathbf{x}_b^{l,t}\!-\!\mathbf{x}_b^{l,*}\|_2^2\!+\!2\tau\eta_b^t\left \langle \mathbf{z}_b^t\!-\!\boldsymbol{\gamma}_b^*, \mathbf{x}_b^{t}\!-\!\mathbf{x}_b^{*}\right \rangle.~~~~~
\end{eqnarray}
According to the Lemma 1 and the step 7 of Algorithm 1, we can put $\mathbf{x}=  \mathbf{z}_b^t=\text{prox}_{\beta\eta_b^t g}(\boldsymbol{\zeta}_b^t-\tau\eta_b^t\mathbf{x}_b^t)$ and $\mathbf{y}=  \boldsymbol{\gamma}_b^*=\text{prox}_{\beta\eta_b^t g}(\boldsymbol{\zeta}_b^*-\tau\eta_b^t\mathbf{x}_b^*)$ into \eqref{core}. Then the
first term of LHS of \eqref{c2} can be bounded as follows
\begin{align}\label{bound}
&\|\mathbf{z}_b^t-\boldsymbol{\gamma}_b^*\|_2^2 \leq \|\boldsymbol{\zeta}_b^t-\tau\eta_b^t\mathbf{x}_b^t-(\boldsymbol{\zeta}_b^*-\tau\eta_b^t\mathbf{x}_b^*)\|_2^2\nonumber\\
&-\left(1+\frac{1}{\beta\eta_b^t\mathcal{L}_g}\right)\|\boldsymbol{\zeta}_b^t-\tau\eta_b^t\mathbf{x}_b^t-\mathbf{z}_b^t-(\boldsymbol{\zeta}_b^*-\tau\eta_b^t\mathbf{x}_b^*-\boldsymbol{\gamma}_b^*)\|_2^2\nonumber\\
&\leq -2\tau\eta_b^t\left \langle\boldsymbol{\zeta}_b^t-\boldsymbol{\zeta}_b^*, \mathbf{x}_b^t-\mathbf{x}_b^*\right \rangle
- \|\boldsymbol{\zeta}_b^t-\mathbf{z}_b^t-(\boldsymbol{\zeta}_b^*-\boldsymbol{\gamma}_b^*)\|_2^2
\nonumber\\
&+ 2\tau\eta_b^t \left \langle\boldsymbol{\zeta}_b^t-\mathbf{z}_b^t-(\boldsymbol{\zeta}_b^*-\boldsymbol{\gamma}_b^*), \mathbf{x}_b^t-\mathbf{x}_b^*\right \rangle+\|\boldsymbol{\zeta}_b^t-\boldsymbol{\zeta}_b^*\|_2^2\nonumber\\
&=\|\boldsymbol{\zeta}_b^t-\boldsymbol{\zeta}_b^*\|_2^2-\|\boldsymbol{\zeta}_b^t-\mathbf{z}_b^t-(\boldsymbol{\zeta}_b^*-\boldsymbol{\gamma}_b^*)\|_2^2\nonumber\\
&-2\tau\eta_b^t \left \langle\mathbf{z}_b^t-\boldsymbol{\gamma}_b^*, \mathbf{x}_b^t-\mathbf{x}_b^*\right \rangle,
\end{align}
where $\mathcal{L}_g$ denotes the smoothness constant of sparsity-promoting function, the second inequality is obtained by omitting the non-positive term multiplied by the factor $1/(\beta\eta_b^t\mathcal{L}_g)$, and the RHS of the last equation is obtained by amalgamating the second and the third terms of its left hand. Combining the inequality in \eqref{bound}, we obtain the upper bound for \eqref{c2}
in a single expression as follows:
\begin{eqnarray}\label{c3}
\!\!\!\!\!\!&&\!\!\!\!\!\!\mathbb{E}\left \{\|\mathbf{z}_b^t+\tau\eta_b^{l,t} \mathbf{x}_b^{l,t}-(\boldsymbol{\gamma}_b^*+\tau\eta_b^{l,t} \mathbf{x}_b^{l,*})\|_2^2\right\}\leq\|\boldsymbol{\zeta}_b^t-\boldsymbol{\zeta}_b^*\|_2^2\nonumber\\
\!\!\!\!\!\!&&\!\!\!\!\!\!\!-\! \|\boldsymbol{\zeta}_b^t\!-\!\mathbf{z}_b^t\!-\!(\boldsymbol{\zeta}_b^*\!-\!\boldsymbol{\gamma}_b^*)\|_2^2\!+\!(\tau\eta_b^t)^2\sum_{l\in \mathcal{N}_b}\frac{(c_{lb}^t)^2\|\mathbf{x}_b^{l,t}\!-\!\mathbf{x}_b^{l,*}\|_2^2}{p_l}.~~~~~
\end{eqnarray}
To obtain the bound of the second term of $\mathcal{W}_b^t$, we add the following missing sum at the both sides of the inequality based on \eqref{c1y} and \eqref{c1}.
\begin{align}
&\label{cc3f} \mathbb{E}\left \{ \sum_{l\in \mathcal{N}_b}(\tau\eta_b^{l,t+1})^2\|\mathbf{x}_b^{l,t+1}-\mathbf{x}_b^{l,*}\|_2^2 \right \} \\
&=\mathbb{E}\left \{(\tau\eta_b^{l,t+1})^2 \|\mathbf{x}_b^{l,t+1}-\mathbf{x}_b^{l,*}\|_2^2\right\}\nonumber\\
&+\mathbb{E}\left \{ \sum_{k\in \mathcal{N}_b,k\neq l}(\tau\eta_b^{k,t+1})^2\|\mathbf{x}_b^{k,t+1}-\mathbf{x}_b^{k,*}\|_2^2 \right \}\nonumber\\
&=\mathbb{E}\left\{(\tau\eta_b^{l,t+1})^2 \|\mathbf{x}_b^{l,t+1}-\mathbf{x}_b^{l,*}\|_2^2\right\}\nonumber\\
&+\mathbb{E}\left \{ \sum_{k\in \mathcal{N}_b,k\neq l}(\tau\eta_b^{l,t})^2\|\mathbf{x}_b^{k,t}-\mathbf{x}_b^{k,*}\|_2^2 \right \}\nonumber\\
&=\mathbb{E}\left \{(\tau\eta_b^{l,t+1})^2 \|\mathbf{x}_b^{l,t+1}-\mathbf{x}_b^{l,*}\|_2^2\right\}\nonumber\\
&+\sum_{l\in \mathcal{N}_b}(1-p_l)(\tau\eta_b^{l,t})^2\|\mathbf{x}_b^{l,t}-\mathbf{x}_b^{l,*}\|_2^2\nonumber\\
&=\sum_{l\in \mathcal{N}_b}(\tau\eta_b^{l,t})^2\|\mathbf{x}_b^{l,t}-\mathbf{x}_b^{l,*}\|_2^2-(\tau\eta_b^t)^2\sum_{l\in \mathcal{N}_b}\frac{(c_{lb}^t)^2}{p_l}\|\mathbf{x}_b^{l,t}-\mathbf{x}_b^{l,*}\|_2^2\nonumber\\
&+\mathbb{E}\left \{(\tau\eta_b^{l,t+1})^2 \|\mathbf{x}_b^{l,t+1}-\mathbf{x}_b^{l,*}\|_2^2\right\}, \label{cc3}
\end{align}
where the second equality stems from the fact that the second term of the RHS of the first equality does not change at iteration $t$, and the last equality is obtained by incorporating the definition of $\eta_b^{l,t}$. Adding \eqref{cc3f} and \eqref{cc3} to \eqref{c1y} and \eqref{c1}, respectively. Then, combining \eqref{c3}, it is found that the last term of RHS of \eqref{c3} vanishes because of the same in \eqref{cc3}. We arrive at
\begin{align}\label{bound1}
&\sum_{l\in \mathcal{N}_b}(\tau\eta_b^{l,t+1})^2\mathbb{E}\left \{\|\mathbf{x}_b^{l,t+1}-\mathbf{x}_b^{l,*}\|_2^2\right\}+\mathbb{E}\left \{\|\boldsymbol{\gamma}_b^{t+1}-\boldsymbol{\gamma}_b^*\|_2^2\right\}\nonumber\\
&\leq\sum_{l\in \mathcal{N}_b}(\tau\eta_b^{l,t})^2\mathbb{E}\left \{\|\mathbf{x}_b^{l,t}-\mathbf{x}_b^{l,*}\|_2^2\right\}-\frac{(\tau\eta_b^{l,t})^2 \|\mathbf{x}_b^{l,t+1}-\mathbf{x}_b^{l,*}\|_2^2}{\tau\eta_b^{l,t}\mathcal{L}_{\Psi_l}}\nonumber\\
&+\mathbb{E}\left \{\|\boldsymbol{\zeta}_b^t-\boldsymbol{\zeta}_b^*\|_2^2\right\}-\mathbb{E}\left \{\|\boldsymbol{\zeta}_b^t-\mathbf{z}_b^t-(\boldsymbol{\zeta}_b^*-\boldsymbol{\gamma}_b^*)\|_2^2\right\}\nonumber\\
&\leq\underbrace{\mathbb{E}\left \{\left \| \boldsymbol{\gamma}_b^t-\boldsymbol{\gamma}_b^* \right \|_2^2\right\}+\sum_{l\in \mathcal{N}_b}(\tau\eta_b^{l,t})^2\mathbb{E}\left \{\|\mathbf{x}_b^{l,t}-\mathbf{x}_b^{l,*}\|_2^2\right\}}_{{\mathcal{W}}_b^{t}}\nonumber\\
&-\bar{p}\varpi \eta_{b}^td_f(\boldsymbol{\gamma}_b^t,\boldsymbol{\gamma}_b^*)-\frac{(\tau\eta_b^{l,t})^2 \|\mathbf{x}_b^{l,t+1}-\mathbf{x}_b^{l,*})\|_2^2}{\tau\eta_b^{l,t}\mathcal{L}_{\Psi_l}}\nonumber\\
&-\mathbb{E}\left \{\|\boldsymbol{\zeta}_b^t-\mathbf{z}_b^t-(\boldsymbol{\zeta}_b^*-\boldsymbol{\gamma}_b^*)\|_2^2\right\},
\end{align}
where the last inequality follows from the fact that, as seen in
the Lemma 1, the difference between $\boldsymbol{\zeta}_b^t$ and $\boldsymbol{\zeta}_b^*$ has an upper bound if the step size satisfies the condition in \eqref{ss}. Omitting the last two terms of the expression \eqref{bound1} leads to
\begin{eqnarray}\label{simf}
{\mathcal{W}}_b^{t+1}\leq{\mathcal{W}}_b^{t} -\bar{p}\varpi \eta_{b}^td_f(\boldsymbol{\gamma}_b^t,\boldsymbol{\gamma}_b^*).
\end{eqnarray}
By telescoping the inequality in \eqref{simf} from 0 to $t$, we have
\begin{eqnarray}\label{simf1}
\!\!\!\!\sum_{j=0}^t\! d_f(\boldsymbol{\gamma}_b^t,\boldsymbol{\gamma}_b^*)\!\!\!&\leq&\!\!\! t\mathbb{E}\left \{d_f(\bar{\boldsymbol{\gamma}}_b^t,\boldsymbol{\gamma}_b^*)\right\}\!\leq\!\frac{1}{\bar{p}\varpi \eta_{b}^0}({\mathcal{W}}_b^{0}\!-\!{\mathcal{W}}_b^{t+1}).~~~~~
\end{eqnarray}
Omitting the last term inside the round brackets and then dividing out both sides of \eqref{simf1} by $t$, we obtain
\begin{eqnarray}\label{theo}
\mathbb{E}\left \{d_f(\bar{\boldsymbol{\gamma}}_b^t,\boldsymbol{\gamma}_b^*)\right\}\leq\frac{1}{t\bar{p}\varpi \eta_{b}^0}{\mathcal{W}}_b^{0}.
\end{eqnarray}
In the sequel, collecting above Bregman divergence bound over all APs results in \eqref{tem1}.
\end{IEEEproof}

From \eqref{theo}, we note that the proposed algorithm converges at the scale of $\mathcal{O}(1/t)$ and it clearly shows how the probability $\bar{p}$ affects the convergence rate. When $\bar{p}=1$, i.e., the case for exact the full gradient calculation, the Bregman divergence bound reaches the minimum value. In addition, the sparsity parameter $\beta$ has vanished in inequality \eqref{bound}, while the similarity parameter $\tau$ contained in ${\mathcal{W}}_b^{0}$ of \eqref{theo} does not equal to $0$, which means that a small value of similarity parameter $\tau$ is required for keeping admissible Bregman divergence bound. The optimum parameters $\beta$ and $\tau$ can be designed based on the steady state
mean-square deviation (MSD), i.e., $\mathbb{E}(\|\text{col}\{\boldsymbol{\gamma}_1^*-{\boldsymbol{\gamma}}_1^t,\boldsymbol{\gamma}_2^*-{\boldsymbol{\gamma}}_2^t,\cdots,\boldsymbol{\gamma}_{B}^*-{\boldsymbol{\gamma}}_B^t\}\|_2^2)$
with $t\rightarrow \infty$, which will be discussed in the future work.

\section{Numerical Results}
In this section, we present extensive simulation results to validate the effectiveness of the proposed cooperative sourced and unsourced random access schemes in 6G wireless networks. We simulate the 6G cell-free wireless network comprising $B=20$ APs geographically distributed in a vast area to serve $N$ potential devices. The AP-to-AP distance is $0.5$ km and the radius of the network coverage is set as $1.8$ km. The positive constants $\theta$ is selected to be $1/0.039$. The penalty parameters $\beta$ and $\tau$ are set as $0.38$ and $0.03$, respectively. $p_l$ is set to $\frac{1}{|\mathcal{N}_b|_c}$, and the constant $\rho$ is set to $500$.

\subsection{Activity Detection for Cooperative Sourced Random Access}
We first conduct simulations to validate the effectiveness of the proposed CAD algorithm for device activity detection. As a performance measure, we adopt the activity error rate (AER). The AER is a sum of the missed detection probability, defined as the probability that a device is active but is declared to be inactive, and the false-alarm probability, defined as the probability that a device is inactive but the detector declares it to be active. As a reference, we compare the proposed CAD algorithm with two baseline schemes: the conventional ML-based multi-cell massive MIMO \cite{covar} and the AMP-based multi-cell massive MIMO \cite{multicell}, where each AP only serves its cell's devices without multi-cell cooperation and treats the inter-cell interference as noise. The average signal-to-noise ratio (SNR) of generic AP $b$ and active device $k\in \mathcal{K}$ over $L$ sequence dimensions is given by SNR$_{b,k}=\frac{\|\mathbf{s}_k\|_2^2g_{b,k}\mathbb{E}\left\{\|\mathbf{h}_{b,k}\|_2^2\right\}}{\mathbb{E}\left\{\|\mathbf{W}_b\|_F^2\right\}}=\frac{g_{b,k}}{\sigma^2}$.

\begin{figure}[h]
  \centering
\includegraphics [width=3in] {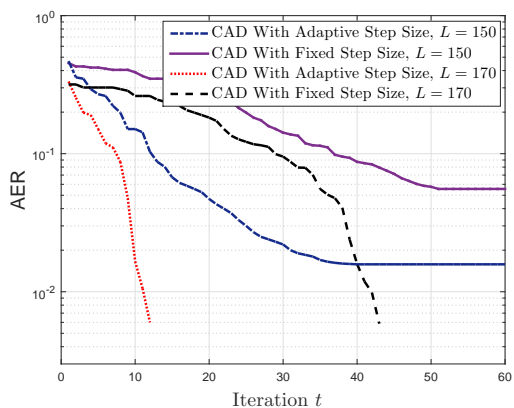}
\caption{The AER for different step sizes with $N = 1,000$,
 $K=300$, $M = 32$, and SNR = $10$ dB.}
\label{stepsize}
\end{figure}

We first validate the analytic convergence rate analysis by numerically comparing the detection performance of CAD, ML-based, and AMP-based multi-cell methods in Fig. \ref{stepsize}. Herein, in order to study the influence of the step size separately, we fix $\bar{p}$ in the proposed CAD algorithm to 1. The adjustment parameter $\epsilon$ is set to $90$, the number of APs for cooperation is set to $5$, and the step size $\eta_b=0.03$ for the CAD algorithm with a fixed step size which is identical for all the APs. The simulation result shows that the proposed CAD with the adaptive step size converges much faster than that of the proposed CAD algorithm with the fixed step size. This is due to the fact that the CAD algorithm with adaptive step size can always choose a more appropriate step size, which allows the algorithm to converge fast at the very first few iterations and quickly attains admissible results in terms of the relatively small generalization error.

\begin{figure}[h]
  \centering
\includegraphics [width=3in] {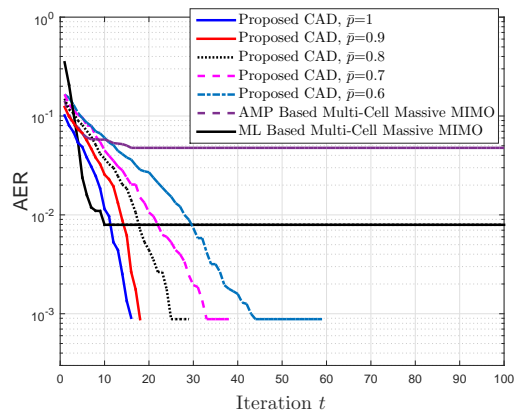}
\caption{The AER for different probabilities $\bar{p}$ with $N = 1,000$,
$K=200$, $M = 32$, $L=120$, and SNR = $10$ dB.}
\label{con_pro}
\end{figure}

Next, we study the impact of $\bar{p}$ on the performance of the CAD algorithm. The parameter $\epsilon$ is set to $90$ and the number of APs for cooperation is set to $5$ in Fig. \ref{con_pro}. The results indicate that for relatively large $\bar{p}$, it slightly influences the convergence rate and the detection accuracy of the proposed scheme. Although the effect is obvious for smaller $\bar{p}$, the case still obtains a huge performance gain compared with the ML-based and the AMP-based multi-cell algorithms in terms of AER. It is observed that the ML-based and the AMP-based multi-cell algorithms all converge more rapidly at the beginning, but reach a plateau afterward and their performance can not be further improved with the number of iterations. Agreeing with the discussions in Section V. B, CAD algorithm enjoys small computational complexity for $0<\bar{p}<1$, since fewer gradients need to be calculated in each iteration. Our results also indicate that it should be possible to throw away some gradient calculations while keeping the almost the same convergence rate and detection accuracies.

\begin{figure}[h]
  \centering
\includegraphics [width=3in] {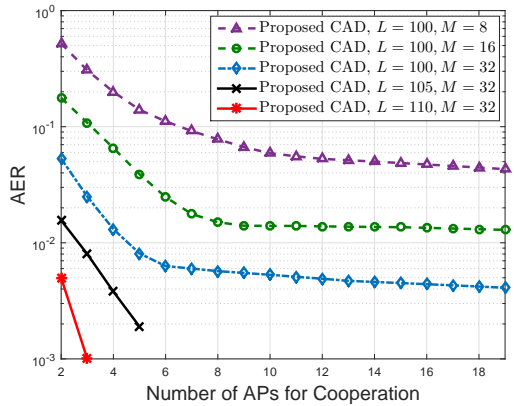}
\caption{The AER for different numbers of APs for cooperation with $N = 1,000$, $K=200$, and SNR = $10$ dB.}
\label{edgedegree}
\end{figure}

In the rest of the simulations, the probability of gradient calculation $\bar{p}$ is set to 1 for unveiling the full potential of CAD under various system settings. Fig. \ref{edgedegree} depicts how detection performance of the proposed CAD algorithm changes with different choices of the number of APs for cooperation. Initially, in the regime with a few numbers of cooperation APs, the AER decreases sharply as the number of cooperation APs increases. However, when the number of cooperation APs continues to increase, the performance improvement diminishes. In addition, it is observed that such a performance saturation point value depends on the number of AP antennas $M$ and the pilot sequence length $L$. Specifically, increasing $M$ or $L$ helps lower the saturation point value, which indicates that the AP becomes more capable of detecting the activity with a low communication cost. The reason for this phenomenon is that for an arbitrary AP, more cooperative connections result in more intermediate estimate exchanges in the proposed CAD algorithm, leading to the improved detection performance. However, the channel strengths from a specific active device to far away APs are negligible, and the exchange of intermediate estimates with the remote AP can not further improve the massive detection performance. The results also indicate that only a small number of APs are required for effective cooperation which strikes a tradeoff between the detection performance and the communication cost.

\begin{figure}[h]
  \centering
\includegraphics [width=3in] {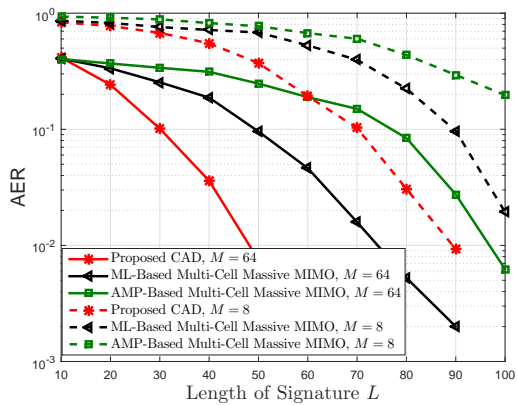}
\caption{The AER for different lengths of sequence $L$ with $N = 500$, $K=100$, and SNR = $10$ dB.}
\label{pilot}
\end{figure}

In the rest of the simulations, the number of APs for cooperation is set to $5$ for illustration. Fig. \ref{pilot} shows the detection performance versus the length of the sequence $L$. From this figure, we can intuitively observe that the AER of all the considered algorithms decrease as the sequence length or the number of antennas of each AP increases. We can also see that the proposed CAD algorithm reduces the required length of sequences for achieving an accurate device activity detection compared with both the ML-based and the AMP-based multi-cell algorithms. Such an advantage of cooperative strategies mainly comes from the fact that the proposed algorithm exploits the structured sparsity and similarity of the multiple APs. Besides, the closed-form expressions for the proximal operator of similarity-promoting and sparsity-promoting terms are derived to achieve higher efficiency. In contrast, the ML-based and the AMP-based multi-cell approaches ignore such prior information and only perform activity detection for the devices distributed in its own cell.

\begin{figure}[h]
  \centering
\includegraphics [width=3in] {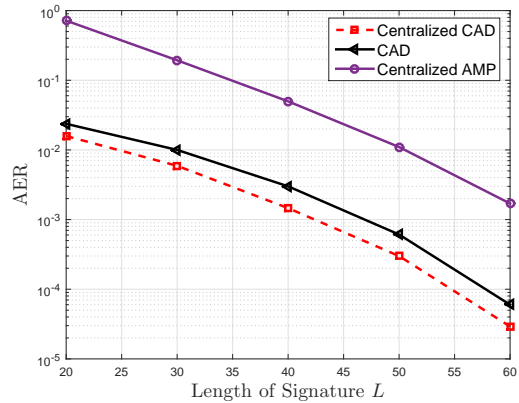}
\caption{Comparison of the proposed CAD algorithm and centralized algorithms with $N = 1,000$, $K=100$, $M=64$ and SNR = $10$ dB.}
\label{centralized}
\end{figure}

Then, we compare the proposed CAD algorithm with the corresponding centralized CAD algorithm and the centralized AMP algorithm. Note that when all APs are connected to a centralized centre and the signals received at all APs are centrally processed, the information from all APs is shared. In this context, the CAD algorithm reduce to the centralized CAD algorithm since $|\mathcal{N}_b|_c=N$ is considered. Moreover, applying the AMP algorithm to jointly process the signals received at all APs can lead to the centralized AMP algorithm. In Fig. \ref{centralized}, we show the AER curves of these algorithms. It is observed that the CAD algorithm has some inevitable performance loss compared with the centralized CAD algorithm due to the less shared information. Note that although the centralized AMP algorithm can process the signals received at all APs, it needs to estimate a larger number of unknown parameters compared with the proposed CAD algorithm, which introduces an unnecessary challenge to the estimation. On the other hand, the centralized AMP algorithm does not consider the similarities among the device state vectors. Thus, the proposed CAD is more efficient for activity detection than that of the centralized AMP algorithm.

\begin{figure}[h]
  \centering
\includegraphics [width=3in] {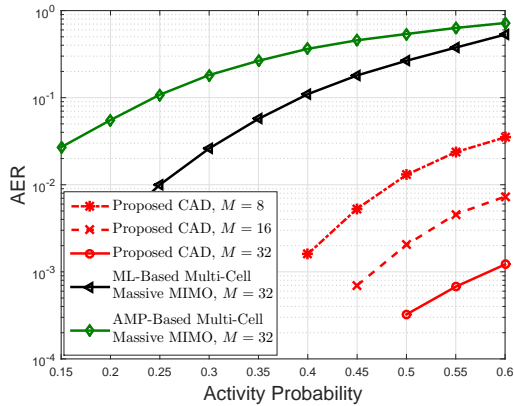}
\caption{The AER for different activity probabilities with $N = 500$, $L=100$, and SNR = $10$ dB.}
\label{sparsity}
\end{figure}

Fig. \ref{sparsity} plots the AER curves of the considered algorithms against the different activity probability. For the comparison of the performance of the proposed scheme and the multi-cell algorithms, we see that the performance of all the algorithms is degraded as the activity probability increases. This is because the co-channel interference among devices increases as more devices are active. Note that the AMP algorithm requires the knowledge of large-scale fading coefficients and the number of active devices $K$, while the CAD algorithm only requires the sample covariance of channel observations. Thus, the proposed CAD algorithm is robust to the inaccurate knowledge of $K$ and the variation in the channel statistics compared with the AMP algorithm. Moreover, the CAD algorithm does not require the explicit knowledge of channel strengths which only needs to estimate a smaller number of unknown parameters, thus, it is more efficient for activity detection than that of the AMP-based multi-cell algorithm. The proposed algorithm outperforms the two multi-cell algorithms by a large margin even if the activity probability is higher than $0.6$. In practice, the proposed algorithm is appealing for the spread applications of IoT with a wide range of activity probability.

\begin{figure}[h]
  \centering
\includegraphics [width=3in] {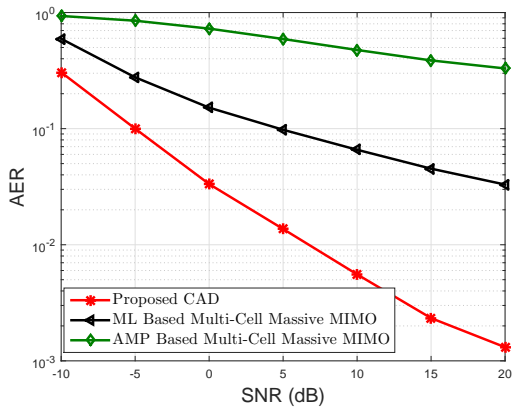}
\caption{The AER for different SNRs with $N = 1,000$, $K=100$, $L=50$, and $M = 32$ antennas at each AP.}
\label{snr}
\end{figure}

Fig. \ref{snr} plots the detection performance with different SNRs. It is observed that for the considered range of SNRs, the ML-based and the AMP-based multi-cell algorithms perform worse than that of the CAD algorithm, and the performance gap is enlarged as the SNR increases. The reason is that the inter-cell interference is a severe limiting factor for reliable activity detection in multi-cell based algorithms, especially when inter-cell interference is dominated in high SNR. While there are no cell boundaries in 6G cell-free wireless networks and the inter-cell interference can be avoided via the APs' cooperation.

\subsection{Data Detection for Cooperative Unsourced Random Access}
For the proposed cooperative massive access scheme designed for cell-free massive MIMO based IoT, we further compare the data detection for the unsourced random access. Note that the probability of error adopted here is different from the definition in sourced random access. The system performance is expressed in terms of probability of missed detection of the per device and per AP, defined as the average fraction of transmitted messages not contained in the list, i.e., $P_{b}^{\mathrm{MD}}=\frac{1}{K}\sum_{k\in\mathcal{K}}\mathbb{P}(m_{b,k}\notin \mathcal{L}_b)$, and the probability of false-alarm, defined as the average fraction of decoded messages that were indeed not sent, i.e., $P_{b}^{\mathrm{FA}}=\frac{|\mathcal{L}_b\backslash \{m_{b,k}:k\in \mathcal{K}\}|_c}{|\mathcal{L}_b|_c}$. We use the probability of error $P_e=\sum_{b=1}^B(P_{b}^{\mathrm{MD}}+P_{b}^{\mathrm{FA}})/B$ to measure the system detection performance. As a reference, we compare the proposed approach with the ML-based multi-cell non-cooperative activity detection for unsourced random access under the same spectral efficiency \cite{unsourced}.

\begin{figure}[h]
  \centering
\includegraphics [width=3in] {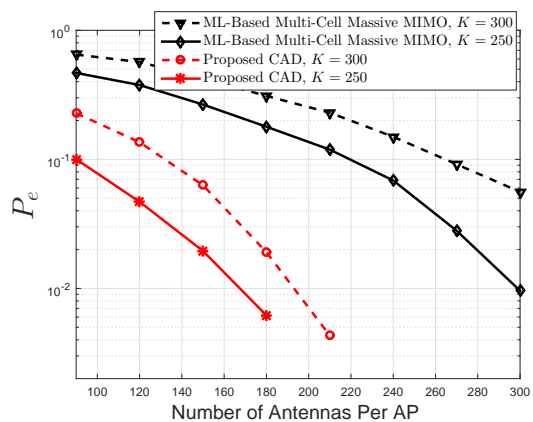}
\caption{Error probability $P_e$ for different $M$ with $J = 12$, $\bar{b}=96$, $\mathcal{Z}=32$, $L=100$, $u=3,200$, and transmit power = $0$ dB.}
\label{unsourced_antenna}
\end{figure}

In Fig. \ref{unsourced_antenna}, we set the energy per bit, $E_b$, and per component AWGN variance $\sigma^2$ such that transmit power $E_b/\sigma^2=0$ dB. Fig. \ref{unsourced_antenna} confirms that as the number of antennas at each AP, $M$, increases, the detection error of the proposed CAD algorithm drops rapidly and faster than that of the ML-based multi-cell algorithm, indicating that employing cooperative AP can quickly drive the detection error to zero with a fewer number of antennas for saving cost. We note that when the number of antennas at each AP exceeds a certain point, the error probability in data detection vanishes with the proposed CAD algorithm. In other words, the superiority of the proposed CAD algorithm is evident in massive MIMO systems, which is a key technique for 6G wireless networks.

\begin{figure}[h]
  \centering
\includegraphics [width=3in] {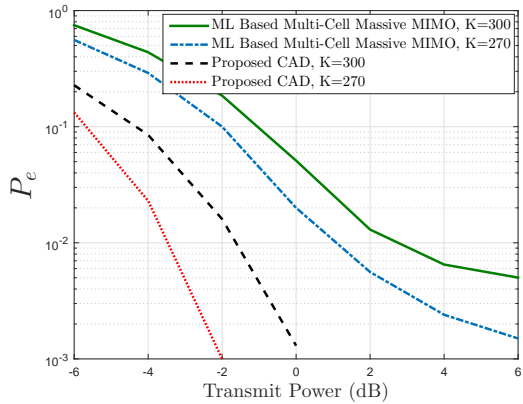}
\caption{Error probability $P_e$ for different transmit powers with $J = 12$, $\bar{b}=96$, $\mathcal{Z}=32$, $L=100$, $u=3,200$, and $M=300$.}
\label{EBN0}
\end{figure}

Fig. \ref{EBN0} plots the error probability versus the transmit power. As expected, increasing the transmit power improves the performance appreciably and
the CAD algorithm achieves much better performance than that of the ML-based multi-cell method. It is also seen that the ML-based multi-cell method is not as efficient as the proposed  cooperative detection method in accommodating more active devices $K$, which is a necessary requirement for practical IoT applications. The simulation result clearly illustrates the advantage of cooperative strategies which stems from that the proposed scheme not only exploits the covariance of local received signal, but also explores the jointly sparse structure and similarity of device state vectors among different APs to enhance the detection performance.
\section{Conclusion}
This paper designed a grant-free cooperative random access framework for sourced and unsourced random access in 6G cell-free wireless networks based on the covariance of the received signals. A cooperative strategy was proposed for activity detection which involves cooperation among adjacent APs in order to exploit joint similarity and sparsity. The developed high-accuracy and low-complexity CAD algorithm handles non-differentiable penalty terms and is shown to converge to the optimality for adaptive step-sizes. Finally, simulation results were presented to illustrate the benefit of cooperation and the proposed algorithm for sourced and unsourced random access can both achieve near-optimal activity detection performance.
\begin{appendices}
\section{The Proof of Lemma 1}
Let us elaborate on the second moments of $\boldsymbol{\zeta}_b^t-\boldsymbol{\zeta}_b^*$ for any $\eta_b^t\leq\eta_{b,0}$ for some constant $\eta_{b,0}$.
\begin{align}\label{sec}
&\|\boldsymbol{\zeta}_b^t-\boldsymbol{\zeta}_b^*\|_2^2\!=\!\|\boldsymbol{\gamma}_b^t\!-\!\boldsymbol{\gamma}_b^*\|_2^2\!-\!2\bar{p}\eta_b^t\left \langle \triangledown f(\boldsymbol{\gamma}_b^t)\!-\!\triangledown f(\boldsymbol{\gamma}_b^*),\boldsymbol{\gamma}_b^t\!-\!\boldsymbol{\gamma}_b^* \right \rangle \nonumber\\ &+\bar{p}^2(\eta_b^t)^2\|\triangledown f(\boldsymbol{\gamma}_b^t)-\triangledown f(\boldsymbol{\gamma}_b^*)\|_2^2\nonumber\\
&\leq\!\! \|\boldsymbol{\gamma}_b^t\!-\!\boldsymbol{\gamma}_b^*\|_2^2\!-\!\bar{p}\eta_b^t(2\!-\!\eta_{b,0}\mathcal{L}_f)\left \langle \triangledown f(\boldsymbol{\gamma}_b^t)\!-\!\triangledown f(\boldsymbol{\gamma}_b^*),\boldsymbol{\gamma}_b^t-\boldsymbol{\gamma}_b^* \right \rangle \nonumber\\
&-\bar{p}\eta_b^t\eta_{b,0}\mathcal{L}_f\left\langle \triangledown f(\boldsymbol{\gamma}_b^t)-\triangledown f(\boldsymbol{\gamma}_b^*),\boldsymbol{\gamma}_b^t-\boldsymbol{\gamma}_b^* \right\rangle\nonumber\\
&+\bar{p}^2\eta_b^t\eta_{b,0}\|\triangledown f(\boldsymbol{\gamma}_b^t)-\triangledown f(\boldsymbol{\gamma}_b^*)\|_2^2.
\end{align}
According to the properties of Bregman divergences $d_f(\cdot,\cdot)$ for a convex and $\mathcal{L}_f$-smooth $f(\cdot)$, we have
\begin{eqnarray}\label{pro1}
\!\!\!\!\!\!\!\!&&\!\!\!\!\!\!\!\!\|\triangledown f(\boldsymbol{\gamma}_b^t)\!-\!\triangledown f(\boldsymbol{\gamma}_b^*)\|_2^2\leq \mathcal{L}_f \left\langle \triangledown f(\boldsymbol{\gamma}_b^t)\!-\!\triangledown f(\boldsymbol{\gamma}_b^*),\boldsymbol{\gamma}_b^t\!-\!\boldsymbol{\gamma}_b^* \right\rangle,~~~~~
\end{eqnarray}
and
\begin{eqnarray}\label{pro2}
\left\langle \triangledown f(\boldsymbol{\gamma}_b^t)-\triangledown f(\boldsymbol{\gamma}_b^*),\boldsymbol{\gamma}_b^t-\boldsymbol{\gamma}_b^* \right\rangle\geq d_f(\boldsymbol{\gamma}_b^t,\boldsymbol{\gamma}_b^* ).
\end{eqnarray}
Substituting \eqref{pro1} into \eqref{sec} whose last two terms will vanish
\begin{eqnarray}\label{sec1}
\!\!\!\!\!\!&&\!\!\!\!\!\! \|\boldsymbol{\zeta}_b^t-\boldsymbol{\zeta}_b^*\|_2^2\leq \|\boldsymbol{\gamma}_b^t-\boldsymbol{\gamma}_b^*\|_2^2\nonumber\\
\!\!\!\!\!\!&&\!\!\!\!\!\!-\bar{p}\eta_b^t(2-\eta_{b,0}\mathcal{L}_f)\left \langle \triangledown f(\boldsymbol{\gamma}_b^t)-\triangledown f(\boldsymbol{\gamma}_b^*),\boldsymbol{\gamma}_b^t-\boldsymbol{\gamma}_b^* \right \rangle.~~~
\end{eqnarray}
As a consequence, by substituting \eqref{pro2} into \eqref{sec1}, we have
\begin{eqnarray}\label{sec3}
\!\!\!\!\!\!\!\!\!\!\!&&\!\!\!\!\!\!\!\!\!\!\!\|\boldsymbol{\zeta}_b^t-\boldsymbol{\zeta}_b^*\|_2^2\leq \|\boldsymbol{\gamma}_b^t-\boldsymbol{\gamma}_b^*\|_2^2 -\bar{p}\eta_b^t(2-\eta_{b,0}\mathcal{L}_f)d_f(\boldsymbol{\gamma}_b^t,\boldsymbol{\gamma}_b^*).
\end{eqnarray}
Inspecting expression \eqref{sec3}, we observe that the convergence of this recursion requires $2-\eta_{b,0}\mathcal{L}_f>0$, thereby yielding $\eta_{b,0}<\frac{2}{\mathcal{L}_f}$. According to the definition of $\eta_b^t$ in \eqref{sep} and the definition of smoothness of $f(\cdot)$, we have
\begin{equation}\label{vg}
 \frac{1}{\mathcal{L}_f+\epsilon}<\eta_b^t<\frac{1}{\epsilon}.
\end{equation}
Altogether these results, we conclude Lemma 1.

\section{The Proof of Lemma 2}
From the definition of the proximal operator in \eqref{pr}, it holds that
\begin{eqnarray}\label{zO}
\!\!\!\!\!\!&&\!\!\!\!\!\! \mathbf{z}_b^t=\text{prox}_{\beta\eta_b^t g}(\boldsymbol{\gamma}_b^*-\eta_b^t\widetilde{\bigtriangledown f}(\boldsymbol{\gamma}_b^*)-\tau\eta_b^t\mathbf{x}_b^*)=\mathop\text{argmin}\limits_{\mathbf{y}}\left \{ \beta\eta_b^t g(\mathbf{y})\right.\nonumber\\
 \!\!\!\!\!\!&&\!\!\!\!\!\!\left. ~~~~~~ +\frac{1}{2}\|\mathbf{y}-(\boldsymbol{\gamma}_b^*-\eta_b^t\widetilde{\bigtriangledown f}(\boldsymbol{\gamma}_b^*)-\tau\eta_b^t\mathbf{x}_b^*)\|_2^2 \right\}.
\end{eqnarray}
Since the function $g(\boldsymbol{\gamma}_b)$ defined in \eqref{spas} is convex, we have
\begin{eqnarray}\label{covx}
\mathbf{0} \in \mathbf{z}_b^t-(\boldsymbol{\gamma}_b^*-\eta_b^t\widetilde{\bigtriangledown f}(\boldsymbol{\gamma}_b^*)-\tau\eta_b^t\mathbf{x}_b^*)+\beta\eta_b^t\partial g(\mathbf{z}_b^t).
\end{eqnarray}
Combining \eqref{covx} and \eqref{hold} leads to $\boldsymbol{\gamma}_b^{*}=\mathbf{z}_b^{*}$.  Therefore,
the relation \eqref{hold} implies that equation \eqref{lem2} holds. Similarly, equation \eqref{mainstep} implies
\begin{align}\label{sile}
&\boldsymbol{\gamma}_b^{t}=\text{prox}_{\tau\eta_b^{l,t} \Psi_l}(\mathbf{z}_b^*+\tau\eta_b^{l,t} \mathbf{x}_b^{l,*})\nonumber\\
&=\mathop\text{argmin}\limits_{\mathbf{y}}\left \{ \tau \eta_b^{l,t} \Psi_l(\mathbf{y})+\frac{1}{2}\|\mathbf{y}-(\mathbf{z}_b^*+\tau\eta_b^{l,t} \mathbf{x}_b^{l,*})\|_2^2 \right \}.
\end{align}
There exists a unique solution $\boldsymbol{\gamma}_b^{*}$ for this problem due to the convexity of each $\Psi_l$, i.e.,
\begin{equation}\label{usl}
\mathbf{0} \in \boldsymbol{\gamma}_b^{t}-(\mathbf{z}_b^*+\tau\eta_b^{l,t} \mathbf{x}_b^{l,*})+\tau\eta_b^{l,t}\partial\Psi_l(\mathbf{x}_b^{l,t})).
\end{equation}
Using this bound and the relation $\mathbf{x}_b^{l,*}\in \partial \Psi_l (\boldsymbol{\gamma}_b^*)$, it can be easily verified that $\boldsymbol{\gamma}_b^{t}=\boldsymbol{\gamma}_b^{*}$.
Therefore, there exists a vector $\boldsymbol{\gamma}_b^{*}$ satisfying equation \eqref{lem3}. This completes the proof.
\end{appendices}

\end{document}